\documentclass{article}
\usepackage{dcolumn}
\usepackage{bm}
\usepackage{verbatim}       

\usepackage[dvips]{graphicx}
\usepackage{amsthm}
\usepackage{amssymb}
\usepackage{bm}
\usepackage{amstext}
\usepackage{psfrag}
\usepackage{amsmath}
\usepackage{amsfonts}

\newfont{\bb}{msbm10 at 12pt}

\newcommand{\dd}{{\rm d}}
\newcommand{\bd}{\begin{definition}}                
\newcommand{\ed}{\end{definition}}                  
\newcommand{\bc}{\begin{corollary}}                 
\newcommand{\ec}{\end{corollary}}                   
\newcommand{\bl}{\begin{lemma}}                     
\newcommand{\el}{\end{lemma}}                       
\newcommand{\bp}{\begin{proposition}}            
\newcommand{\ep}{\end{proposition}}                
\newcommand{\bere}{\begin{remark}}                  
\newcommand{\ere}{\end{remark}}                     

\newcommand{\bt}{\begin{theorem}}
\newcommand{\et}{\end{theorem}}

\newcommand{\be}{\begin{equation}}
\newcommand{\ee}{\end{equation}}

\newcommand{\bit}{\begin{itemize}}
\newcommand{\eit}{\end{itemize}}
\newtheorem{theorem}{Theorem}[section]
\newtheorem{corollary}[theorem]{Corollary}
\newtheorem{claim}[theorem]{Claim}
\newtheorem{lemma}[theorem]{Lemma}
\newtheorem{proposition}[theorem]{Proposition}
\theoremstyle{definition}
\newtheorem{definition}[theorem]{Definition}
\theoremstyle{remark}
\newtheorem{remark}[theorem]{Remark}
\newtheorem{example}[theorem]{Example}

\begin{document}
%


\title{The causal ladder and the strength of $K$-causality. II}

\author{E. Minguzzi \footnote{Department of Applied Mathematics, Florence
 University, Via S. Marta 3,  50139 Florence, Italy. E-mail: ettore.minguzzi@unifi.it}}

\date{}
\maketitle

\begin{abstract}
\noindent Hawking's stable causality implies Sorkin and Woolgar's
$K$-causality. The work investigates the possible equivalence
between the two causality requirements, an issue which was first
considered by H. Seifert and then raised again by R. Low after the
introduction of $K$-causality. First, a new proof is given that a
spacetime is stably causal iff the Seifert causal relation is a
partial order. It is then shown that given a $K$-causal spacetime
and chosen an event, the light cones can be widened in a
neighborhood of the event without spoiling $K$-causality. The idea
is that this widening of the light cones can be continued leading to
a global one. Unfortunately, due to some difficulties in the
inductive process the author was not able to complete the program
for a proof as originally conceived by H. Seifert. Nevertheless, it
is proved that if $K$-causality coincides with stable causality then
in any $K$-causal spacetime the $K^+$ future coincides with the
Seifert future. Explicit examples are provided which show that the
$K^+$ future may differ from the Seifert future in causal
spacetimes.

\end{abstract}




\section{Introduction}
The property of $K$-causality was introduced about ten years ago by
R.D. Sorkin and E. Woolgar \cite{sorkin96}. Given a spacetime
$(M,g)$ they defined the relation $K^{+}$ as  the smallest closed
subset $K^{+}\subset M\times M$, which contains $I^{+}$,
$I^{+}\subset K^{+}$, and shares the transitivity property:
$(x,y)\in K^{+}$ and $(y,z) \in K^{+}$ $\Rightarrow (x,z) \in K^{+}$
(the set of  causal relations satisfying these properties is
non-empty, consider for instance the trivial subset $M\times M$).
This definition raised from the fact that $J^{+}$ while transitive
is not necessarily closed whereas $\bar{J}^{+}$ while closed is not
necessarily transitive.
 They also defined the spacetime to be  {\em
$K$-causal} if the relation $K^{+}$ is a partial order i.e. not only
transitive and reflexive but also antisymmetric, that is, such that,
$(x, z) \in K^{+}$ and $(z, x) \in K^{+}$ $\Rightarrow x=z$.

R. Low pointed out \cite[footnote p. 1990]{sorkin96} that H.
Seifert's causal relation $J^{+}_S\supset J^{+}$ is closed and
transitive, hence $K^{+}\subset J^{+}_S$. It is then natural to ask
whether it is always $K^{+}=J^{+}_S$ and if this is  not the case,
whether it is at least true that $J^{+}_S$ is a partial order
whenever $K^{+}$ is a partial order.  Seifert proved
\cite{seifert71} the transitivity and closure of $J^{+}_S$ and gave
an argument showing that $J^{+}_S$ is a partial order if and only if
the spacetime is stably causal (for a rigorous proof see \cite[Prop.
2.3]{hawking74}  or theorem \ref{pai} below). As a consequence,
since $K^{+}\subset J_S^{+}$, if the spacetime is stably causal then
it is $K$-causal. Moreover, the equality $K^{+}=J^{+}_S$ would imply
that the properties of $K$-causality and stable causality coincide.
On the contrary, the example of a spacetime $K$-causal but
non-stably causal would imply at once that $K^{+}$ does not always
coincide with $J^{+}_S$ and that $K$-causality can be included in
the causal hierarchy of spacetimes \cite{hawking74,minguzzi06c} just
below stable causality.

Seifert himself \cite{seifert71} raised the problem as to whether
$J^{+}_S$ could be regarded as the smallest closed and transitive
causal relation containing $I^{+}$. One of his lemmas \cite[Lemma
2]{seifert71}  actually answers this question in the affirmative
sense provided $K^{+}$ is a partial order.

\begin{claim} \label{c1}
 (Seifert's \cite[Lemma
2]{seifert71}) \\
$J^{+}_S$ is the smallest among the partial orders $P^{+}\subset
M\times M$ not smaller that $J^{+}$ with closed $P^{+}(x)$ and
$P^{-}(x)$ for all $x$ {}[provided such a smallest partial order
exists] \footnote{Text in square brackets by the author. }.
\end{claim}

 Indeed,  we shall see in section
\ref{mja} that the previous claim can be  conveniently rephrased in
the following way

\begin{claim} \label{c2}
  If $(M,g)$ is $K$-causal  then $K^{+}=J^{+}_S$.
\end{claim}

A consequence is that $K$-causality is equivalent to stable
causality. Unfortunately, Seifert's arguments were not rigorous as
they did not take into account the many subtleties of the $K^{+}$
relation.  This lemma was almost never cited in subsequent
literature and some researchers who tried to reproduce it began to
raise some doubts on its validity. It suffices to say that using it
some
 proofs later given by Hawking and Sachs \cite{hawking74} could have
 been
considerably simplified as I will show in the last section. In that
fundamental work Hawking and Sachs preferred to take a path
independent of Seifert's 1971 work and indeed, although they cited
Seifert, they gave a completely new  proof of the equivalence
between stable causality and the antisymmetry of $J^{+}_S$ and
avoided any mention to  the claim above. Over the time the question
raised by Seifert's work was overlooked and only with the
introduction of the $K^{+}$ relation it was rediscovered from a
different perspective. This work is devoted to the study of this
open issue.

The work is organized as follows.

In section \ref{mja} some general results for binary relations on
$M\times M$ are given. The equivalence between claims \ref{c1} and
\ref{c2} is proved here. In this section as well as in the rest of
the work the reader is assumed to be familiar with the conventions
and notations introduced in \cite[Sect. 1 and 2]{minguzzi07b}. Let
me just remind that the spacetime signature is $(-,+,\cdots,+)$,
that the subset inclusion is reflexive, $X\subset X$, and that the
boundary of a set $A$ is denoted $\dot{A}$.

Section \ref{mka2}  deals with Seifert's closed relation. I
generalize, simplify and fill gaps of some proofs, particularly that
on the equivalence between stable causality and the antisymmetry of
$J^{+}_S$, the extent of the improvements being there explained.

Section \ref{vio} deals with some results on the violating sets for
$J^{+}_S$.

In section \ref{cex} some examples of spacetimes in which $K^{+}\ne
J^{+}_S$ are given. The strategy outlined by Seifert for a proof of
the equivalence between $K$-causality and stable causality is, more
or less, followed here and made rigorous. Seifert suggested to prove
that (i) $K$-causality implies strong $K$-causality (\cite[Lemma
16]{sorkin96} and lemma \ref{sor3} below), and that (ii) strong
$K$-causality implies that, chosen an event, the cones can be
widened in a neighborhood of the chosen point while preserving
$K$-causality (theorem \ref{fgh}) (iii) the process of widening the
light cones can be continued so as to obtain a global widening of
the light cones and hence the proof that a $K$-causal spacetime is
stably $K$-causal and hence stably causal (actually stable causality
and stable $K$-causality coincide, see corollary \ref{pxe}).

 The same strategy is followed here
although in several points the proofs differ significatively from
what originally suspected by Seifert. Several technical lemmas are
required  because what is intuitive for $J^{+}$ is usually hard to
prove for $K^{+}$, the reason lying on the fact that $K^{+}$ is
defined through its closure and transitivity properties and not, at
least not straightforwardly, by using the set of causal curves.
Indeed, Sorkin and Woolgar \cite{sorkin96} introduced the method of
transfinite induction in order to obtain some basic results. Here it
is shown that this method can be avoided, and that it can be
replaced by soft topological arguments which take advantage of the
minimality of $K^{+}$.

Unfortunately,  I will not be able to prove step (iii). The process
of widening the light cones can indeed be indefinitely continued but
some technicalities do not allow to conclude that a global widening
of the light cones can be obtained.

Nevertheless, in section \ref{pos} it is proved that if it is true
that $K$-causality coincides with stable causality then it is also
true that in a $K$-causal spacetime $K^{+}=J^{+}_S$ (theorem
\ref{las}). Under the same assumption a very simple proof that
causal continuity implies stable causality is given. In this respect
the proof of the equivalence between $K$-causality and stable
causality is recognized as an important problem in causality theory
as many old proof would be greatly simplified by the knowledge of
this result, and new ones would follow.

Despite the fact that the main problem has not been solved, the
sections \ref{cex} and \ref{pos} contain many new results and
properties of the $K^+$ relation which may prove to be useful in
future applications.

\section{Preliminaries} \label{mja}

The reader is assumed to be familiar with  \cite[Sect.
2]{minguzzi07b} which contains some definitions and results for a
 generic binary relation $R^{+}\subset M\times M$. The definitions of
closure, partial closure, transitivity, reflexivity, antisymmetry,
$R$-causality, strong $R$-causality, stable $R$-causality will not
be repeated here. Nor will be repeated the definitions of the sets
$R^{\pm}(x)$ or $R^{-}$ given $R^{+}$ or of the diagonal $\Delta$ in
$M\times M$.

Here I give some useful definitions  for dealing with violations of
the antisymmety condition. The { \em $R$-causality violating set on}
$M\times M$ is
\begin{equation}
VR=R^{+}\cap R^{-} \cap \Delta^{C},
\end{equation}
which is the set of pairs at which the antisymmetric condition for
$R^{+}$ (and hence $R^{-}$) fails. Note that since $\Delta$ is
closed $VR$ is open whenever $R^{+}$ is open, for instance, $VI$ is
open.

Analogously, it is useful to define the {$R$-causality violating set
on} $M$ as $vR=\pi(VR)$, where $\pi:M\times M \to M$ is the
projection on the first factor (or, equivalently, on the second
factor). It is made of all the points $x$ at which $R$-causality at
$x$ does not hold.
%
%

In general given the spacetime $(M,g)$ the causal relation $R^{+}$
will be related in some way to $(M,g)$, for instance $I^{+}$ is the
set of pairs $(x,z)$ such that there is a timelike curve connecting
$x$ to $z$. In order to stress this spacetime dependence I shall
sometimes write $R^{+}_{(M,g)}$. If we are working in the same
spacetime manifold $M$ but with different metrics I may write
$R^{+}_g$ in place of $R^{+}_{(M,g)}$ in order  to stress the metric
(or most often the conformal structure) dependence. Analogously I
may write sentences like ``($\bar{g}$-)causal curve $\gamma$'' to
stress that $\gamma$ is causal with respect to the metric $\bar{g}$,
or ``($\bar{g}$-)convex  set $V$'' to stress that $V$ is a convex
set with respect to the metric $\bar{g}$. If instead we are dealing
with the same metric but with different open sets $A\subset M$, I
may write $R^{+}_A$ in place of $R^{+}_{(A,g)}$. However, the reader
should be careful because $J^{+}_S$ denotes the Seifert future, not
the causal relation for the spacetime $(S,g)$. In order to avoid
confusion   I will not use the letter $S$ for any open set.

Let me recall \cite[Th. 2.2]{minguzzi07b} that a transitive relation
$R^{+}$ which contains $I^{+}$ is closed iff it has closed
$R^{\pm}(x)$ for all $x$, and that, as a consequence \cite[Cor.
2.5]{minguzzi07b}, $K^{+}$ is the smallest  transitive relation
containing $I^{+}$ such that for every $x \in M$, $K^{+}(x)$ and
$K^{-}(x)$ are closed. Theorem 2.2 of \cite{minguzzi07b} implies

\begin{corollary}
Claims \ref{c1} and \ref{c2} are equivalent.
\end{corollary}

\begin{proof}

Note that the smallest relation, call it $M^{+}$, which is
transitive, contains $I^{+}$, is partially closed (or equivalently
closed because of the first two properties) and antisymmetric may
not exist because though all these properties are preserved under
arbitrary intersections of relations, the set of relations
satisfying the properties may be empty. Provided it is not empty
$M^{+}$ exists and  $K^{+} \subset M^{+}$ because $M^{+}$ is
transitive, closed, and contains $I^{+}$. But then the inclusion
implies that $K^{+}$ is antisymmetric too thus the spacetime is
$K$-causal, and by definition of $M^{+}$, $M^{+} \subset K^{+}$.
Thus if $M^{+}$ exists $M^{+}=K^{+}$ and the spacetime is
$K$-causal. Conversely, if the spacetime is $K$-causal then $M^{+}$
exists and hence $M^{+}=K^{+}$. As a consequence, claims \ref{c1}
and \ref{c2} are equivalent.
\end{proof}

Let $\mathcal{P}$ be a conformal invariant property for a spacetime.
Assume moreover that if $\mathcal{P}$ holds for $(M,g_1)$ then it
holds for every $(M, g_2)$, $g_2 <g_1$. Examples for $\mathcal{P}$
are chronology, causality, distinction, $K$-causality and  stable
causality. For such a property the spacetime $(M,g)$ is said to have
the {\em stable-$\mathcal{P}$ property} or to be {\em stably
$\mathcal{P}$ }if there is $g'>g$ such that $(M,g')$ has the
$\mathcal{P}$ property. It is clear that stable-causality is the
usual stable causality. It is also clear that if $\mathcal{P}_1
\Rightarrow \mathcal{P}_2$ then stable-$\mathcal{P}_1 \Rightarrow$
stable-$\mathcal{P}_2$. A nice result is that the operation of
making a property stable is idempotent, that is

\begin{lemma}
Let $\mathcal{P}$ be a conformal invariant property such that if
$\mathcal{P}$ holds for $(M,g_1)$ then it holds for every $(M,
g_2)$, $g_2 <g_1$. Then the property stable-stable-$\mathcal{P}$
coincides with stable-$\mathcal{P}$.
\end{lemma}

\begin{proof}
The implication $\Rightarrow$ is obvious, for the other direction if
$(M,g)$ is stably $\mathcal{P}$ there is $g'>g$ such that $(M,g')$
has property $\mathcal{P}$. Take $g''$, $g<g''<g'$, then $(M,g'')$
is stably $\mathcal{P}$ and hence $(M,g)$ has property
stable-stable-$\mathcal{P}$. \end{proof}

\begin{corollary} \label{pxe}
Stable causality is equivalent to stable chronology, stable
distinction, stable strong causality, stable $K$-causality and
stable stable causality.
\end{corollary}

\begin{proof}
Stable chronology implies causality, because a closed causal curve
for $(M,g)$ is closed and timelike for $(M,g')$, $g'>g$. Thus
stable-chronology= stable-stable-chronology $\Rightarrow$
stable-causality. Now consider the closed chain of implications
stable-chronology $\Rightarrow$
stable-causality=stable-stable-causality $\Rightarrow$
stable-$K$-causality $\Rightarrow$ stable-strong causality
$\Rightarrow$ stable-distinction $\Rightarrow$ stable-causality
$\Rightarrow$ stable chronology, from which the equivalence of all
these properties follows.
\end{proof}

Stable non-total viciousness is distinct from stable causality (see
remark \ref{nos}).

\section{Seifert's closed relation and stable causality} \label{mka2}

In this section the relationship between stable causality and
Seifert's causal relation $J^{+}_S$ is clarified. Concerning the
topology of stable causality, for some results not considered here
nor cited elsewhere in the paper but still of interest the reader
may consult \cite{aguirre89,racz88} and \cite[Sect. 6.4]{hawking73}.

Given two metrics $g$, $g'$, over $M$, denote as usual $g' >g$ if
every causal vector for $g$ is timelike for $g'$, and $g'\ge g$ if
every causal vector for $g$ is causal for $g'$. In presence of
different metrics, the sets $I^{+}_{g}, J^{+}_{g} \subset M\times
M$, are the chronological and causal sets of $(M,g)$.

\begin{definition}
The set $J^{+}_S\subset M\times M$, defining Seifert's causal
relation on $(M,g)$ is
\[J^{+}_S=\{(x,z): (x,z)\in J^{+}_{g'} \textrm{ for every } g'>g \}
=\bigcap_{g'>g} J^{+}_{g'}.
\]

\end{definition}

Sometimes, in order to avoid confusion, I shall write $J^{+}_{S\,
g}$ in place of $J^{+}_S$ to point out the causal structure to which
$J^{+}_S$ refers to.

\begin{lemma} \label{vfe}
If $\tilde{g}<{g}$ then $\bar{J}^{+}_{\tilde{g}} \subset \Delta \cup
I^{+}_{{g}} $.
\end{lemma}

\begin{proof}
Let $(x,z) \in \bar{J}^{+}_{\tilde{g}}\backslash \Delta$, let
$\sigma_n$ be a sequence of ($\tilde{g}$-)causal curves of endpoints
$x_n$, $z_n$ in $(M,\tilde{g})$. Using a limit curve theorem it
follows the existence of a future directed ($\tilde{g}$-)causal
curve $\sigma^x$ starting from $x$, and a past directed
($\tilde{g}$-)causal curve $\sigma^z$ ending at $z$, and a
subsequence $\sigma_j$ distinguishing both curves. Taken $x' \in
\sigma^x$, $z' \in \sigma^z$ it follows $(x,x')\in
J^{+}_{\tilde{g}}$, $(z',z)\in J^{+}_{\tilde{g}}$ and $(x',z') \in
\bar{J}^{+}_{\tilde{g}}$, or, in terms of the causal relations of
${g}$, $(x,x')\in I^{+}_{{g}}$, $(z',z)\in I^{+}_{{g}}$ and $(x',z')
\in \bar{J}^{+}_{{g}}$, which implies, because $I^{+}_{{g}}$ is
open, $(x,z) \in I^{+}_{{g}}$.
\end{proof}

\begin{lemma} \label{mip}
Equivalent definitions of $J^{+}_{S\, g}$ on the spacetime $(M,g)$
are
\[
J^{+}_{S\, g}= \Delta\cup \bigcap_{g'>g} I^{+}_{g'}= \bigcap_{g'>g}
\bar{J}^{+}_{g'},
\]
moreover, $J^{+}_{S \, g}=\bigcap_{g'>g} J^{+}_{S\, g'}$.
\end{lemma}

\begin{proof}
For the first equality we have only to show that $ \bigcap_{g'>g}
J^{+}_{g'} \subset \Delta \cup \bigcap_{g'>g} I^{+}_{g'}$ the other
inclusion being obvious. For every $\bar{g}>g$, taken $\tilde{g}$
such that $g<\tilde{g}<\bar{g}$, it is $ J^{+}_{\tilde{g}} \subset
\Delta \cup I^{+}_{\bar{g}}$, hence $\bigcap_{g'>g} {J}^{+}_{g'}
\subset \Delta \cup I^{+}_{\bar{g}}$, since $\bar{g}>g$ is arbitrary
the thesis follows.

For the second equality we have only to show that $ \bigcap_{g'>g}
\bar{J}^{+}_{g'} \subset \bigcap_{g'>g} J^{+}_{g'}$ the other
inclusion being obvious. Let $\bar{g}>g$ , taken $\tilde{g}$ such
that $g<\tilde{g}<\bar{g}$, by lemma \ref{vfe}  it is $
\bar{J}^{+}_{\tilde{g}} \subset J^{+}_{\bar{g}}$, thus
$\bigcap_{g'>g} \bar{J}^{+}_{g'} \subset J^{+}_{\bar{g}}$, since
$\bar{g}>g$ is arbitrary the  thesis  follows.

For the last statement note that $J^{+}_{S \, g}= \bigcap_{g'>g}
J^{+}_{g'}=\bigcap_{g'>g} \large( \bigcap_{g''>g'} J^{+}_{g''}
\large)=\bigcap_{g'>g} J^{+}_{S \, g'}$ .
\end{proof}

\begin{theorem}
The relation $J^{+}_S$ is closed, transitive and contains $I^{+}$,
moreover for every $x \in M$, $J_S^{+}(x)$ and $J_S^{-}(x)$ are
closed and contain respectively $I^{+}(x)$ and $I^{-}(x)$.
\end{theorem}

\begin{proof}
The transitivity is obvious because for every $g'>g$, $J^{+}_{g'}$
is transitive. The statements on the inclusions of $I^{+}$,
$I^{+}(x)$ and $I^{-}(x)$ are trivial too. The closure of $J^{+}_S$
is immediate from $J^{+}_S=\bigcap_{g'>g} \bar{J}^{+}_{g'}$. The
closure of $J_S^{-}(x)$, and $J_S^{+}(x)$, follows by taking
$(x_n,z_n)\to(x,z)$, $(x_n,z_n) \in J^{+}_S$,  and by using the
closure of $J^{+}_S$ in the cases $x_n=x$ or $z_n=z$ for the initial
choice of converging sequence.

\end{proof}
A simple consequence is $K^{+} \subset  J^{+}_S$ and hence

\begin{corollary} \label{mkq}
If $J^{+}_S$ is a partial order then the spacetime is $K$-causal.
\end{corollary}

\begin{lemma} \label{vfe2}
If $g<g'$ then $K^{+}_{g} \subset \Delta \cup I^{+}_{g'} $.
\end{lemma}

\begin{proof}
It follows from $K^{+}_{g} \subset J^{+}_{S \, g}=\Delta\cup
\bigcap_{g'>g} I^{+}_{g'}=\bigcap_{g'>g}(\Delta\cup I^{+}_{g'})$.
\end{proof}

\begin{lemma} \label{mip2}
An  equivalent definition of $J^{+}_{S\, g}$ on the spacetime
$(M,g)$ is
\[
J^{+}_{S\, g}= \bigcap_{g'>g} K^{+}_{g'}.
\]
\end{lemma}

\begin{proof}

Since $J^{+}_{g'} \subset K^{+}_{g'}$, $ J^{+}_{S \, g}=
\bigcap_{g'>g}J^{+}_{g'} \subset  \bigcap_{g'>g} K^{+}_{g'}$, but,
since $J^{+}_{S\, g'}$ is closed and transitive, $ K^{+}_{g'}
\subset J^{+}_{S\, g'}$ and $ \bigcap_{g'>g} K^{+}_{g'} \subset
\bigcap_{g'>g}  J^{+}_{S\, g'}=J^{+}_{S \, g}$, from which the
thesis follows.

\end{proof}

\begin{remark}
From the definition of the sets $R^{\pm}(x)$ given the binary
relation $R^{+}$ it follows
\[
J^{\pm}_{S\, g}(x)=\{x\}\cup \bigcap_{g'>g}
I^{\pm}_{g'}(x)=\bigcap_{g'>g} J^{\pm}_{g'}(x)=\bigcap_{g'>g}
K^{\pm}_{g'}(x)=\bigcap_{g'>g} J^{\pm}_{S\, g'}(x).
\]
\end{remark}

\begin{lemma} \label{nhu}
If $J^{+}_S$ on $(M,g)$ is a partial order then for every $x \in M$
there is a ($x$-dependent) metric $g_x>g$ such that $(M,g_x)$ is
chronological at $x$.
\end{lemma}

\begin{proof}
Assume, by contradiction, that the thesis does not hold, then there
is $x \in M$ such that  for every $g'>g$, there is a closed
($g'$-)timelike curve $\sigma_{g'}$ passing through $x$.

Let $\bar{g}>g$, introduce a Riemannian metric in a neighborhood of
$x$ and consider $S=\dot{B}(x,\epsilon)$, i.e. the surface of the
ball of Riemannian radius $\epsilon>0$. Choose $\epsilon$
sufficiently small so that $S$ is contained in a ($\bar{g}$-)convex
neighborhood contained in a ($\bar{g}$-)globally hyperbolic
neighborhood $V$.

For every $g'$, $g<g'<\bar{g}$, there is a closed ($g'$-)timelike
curve $\sigma_{g'}$ passing through $x$. This curve must escape the
hyperbolic neighborhood $V$ otherwise in $(V,\bar{g})$ there would
be a closed ($\bar{g}$-)timelike curve. Hence the curve must meet
$S$ at some point of $S\cap J^{+}_{\bar{g}}(x)$. Given $g'$ the
event $x$ belongs to the {\em chronologically violating set} $vI_g'$
which is open \cite{penrose72} and which can be written as the union
of disjoint open sets of the form $I^{+}_{g'}(y)\cap I^{-}_{g'}(y)$
where $y$ is any point of the component \cite[Prop.
4.27]{penrose72}. In particular $x$ belongs to the component
$I^{+}_{g'}(x)\cap I^{-}_{g'}(x)$. The set $A(g')= I^{+}_{g'}(x)\cap
I^{-}_{g'}(x)\cap S \cap I^{+}_{\bar{g}}(x) \ne \emptyset$ is open
in the topology inherited by $S$ and non-empty because $\sigma_{g'}$
must meet $S\cap I^{+}_{\bar{g}}(x)$. In the topology of $S$,
$\bar{A}(g')$ are compact and $\bigcap_{g<g'<\bar{g}}  \bar{A}(g')
\ne \emptyset$.

This  result follows because otherwise $\bigcup_{g<g'<\bar{g}}
\bar{A}^{C}(g') =S $ where the complement $C$ is taken in the
topological space $S$. Since $\bar{A}^{C}(g')$ are open sets there
would be a finite covering $S= \bar{A}^{C}(g_1) \cup \cdots \cup
\bar{A}^{C}(g_k)$. Now, note that if $\hat{g}<\tilde{g}$ then, since
a timelike curve for $\hat{g}$ is timelike for $\tilde{g}$ it is
$A(\hat{g})\subset A(\tilde{g})$ and $ \bar{A}^{C}(\hat{g})\supset
\bar{A}^{C}(\tilde{g})$. A metric $\check{g}$ exists such that for
$i=1,\ldots, k$, $g<\check{g}<g_i$, thus  $ \bar{A}^{C}(\check{g})
\supset \bigcup_{i=1}^k \bar{A}^{C}({g}_i)=S$, hence
$A(\check{g})=\emptyset$ a contradiction.

Thus there is $z \in \bigcap_{g<g'<\bar{g}} \bar{A}(g') \ne
\emptyset $. In other words there is an event $z \in S$ such that
for every $g'>g$ there are  ($g'$-)timelike curves starting from $x$
and passing arbitrarily close to $z$. Thus for every $g'>g$, $(x,z)
\in \bar{J}^{+}_{g'}$ and $(z,x) \in  \bar{J}^{+}_{g'}$ thus by
lemma \ref{mip}, $(x,z) \in J^{+}_S$ and $(z,x) \in J^{+}_S$ but $x
\ne z$, i.e. $J^{+}_S$ is not a partial order.
\end{proof}

\begin{lemma} \label{pag}
If $(M,g)$ is chronological at $x$ then for every $g'<g$, $(M,g')$
is strongly causal at $x$. (Stated in another way, if $(M,g')$ is
non-strongly causal at $x$ then for every $g>g'$, there is a
($g$-)timelike closed curve passing through $x$.)
\end{lemma}

\begin{proof}
If $(M,g')$ is not strongly causal at $x$ then the characterizing
property (ii) of \cite[Lemma 3.22]{minguzzi06c} does not hold, that
is, there is  a neighborhood $U \ni x $ and a sequence of
($g'$-)causal curves $\sigma_n$ of endpoints $x_n,z_n$, with $x_n\to
x$, $z_n \to x$, not entirely contained in $U$. Let $C\ni x$ be a
($g'$-)convex neighborhood whose compact closure is contained in
another ($g'$-)convex neighborhood  $V \subset U$ contained in a
globally hyperbolic neighborhood. Let $c_n\in \dot{C}$ be the first
point at which $\sigma_n$ escapes $C$, and let $d_n$ be the last
point at which $\sigma_n$ reenters $C$. Since $\dot{C}$ is compact
there are $c,d \in \dot C$, and  a subsequence $\sigma_k$ such that
$c_k \to c$, $d_k \to d$ and since $V$ is convex, the causal
relation on $V\times V$, $J^{+}_{(V,g')}$, is closed and hence
$(x,c), (d,x) \in J_{(V,g')}^{+}$ thus $(x,c),(d,x) \in J_{g'}^{+}$
(note that $(d,c)\in J_{(V,g')}^{+}$ thus $d$ and $c$ must be
distinct since the spacetime $(V,g')$ is causal). Taking into
account that $(c_k,d_k) \in J_{g'}^{+}$ it is $(c,d) \in
\bar{J}_{g'}^{+}$. Thus there is a ($g$-)timelike curve connecting
$d$ to $c$ passing through $x$, and since $(c,d) \in
\bar{J}_{g}^{+}$ and $I^{+}_g$ is open there is a closed timelike
curve passing through $x$.
\end{proof}

\begin{figure}
\centering \psfrag{g}{$g$} \psfrag{g1}{$g_1$} \psfrag{g2}{$g_2$}
\psfrag{g3}{$g_3$} \psfrag{g4}{$g_4$} \psfrag{gp}{$g'$}
\psfrag{p}{$\!p$} \psfrag{C1}{$C_1$}
\psfrag{C2}{$C_2$}\psfrag{C3}{$C_3$} \psfrag{A1}{$A_1$}
\psfrag{A2}{$A_2$}\psfrag{A3}{$A_3$} \psfrag{B1}{$B_1$}
\psfrag{B2}{$B_2$}\psfrag{B3}{$B_3$} \psfrag{F}{$\gamma$}
\psfrag{M}{$M$}
\includegraphics[width=8cm]{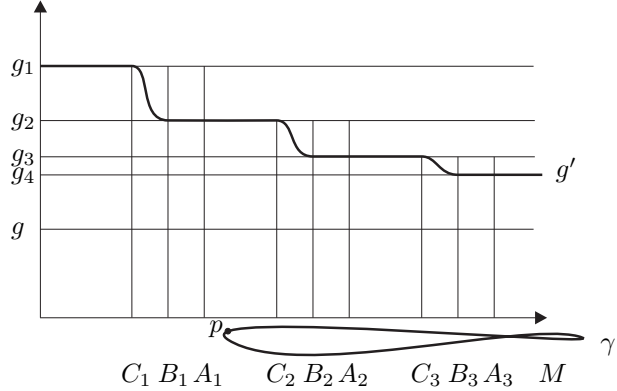}
\caption{The idea behind the proof of lemma \ref{vax}. If the event
$x$ of the statement does not exist there is a sequence of compacts
$C_n$ and metrics $g_n>g$ such that $(M,g_n)$ is chronological in
$A_n\supset C_n$. Then a metric $g'>g$  exists which is
chronological everywhere (in contradiction with the non-stable
causality of $(M,g)$), indeed, if not, a closed timelike curve
$\gamma$ would have a point $p\in \bar{B}_i$ with $i$ lowest
possible index, then $(M,g_i)$ would not be chronological at $p \in
A_i$ (in the figure $i=2$) a contradiction.} \label{dim}
\end{figure}

\begin{lemma} \label{vax}
If for every $x \in M$ there is a ($x$ dependent) $g_x>g$ such that
$(M,g_x)$ is chronological at $x$ then $(M,g)$ is stably causal.
(Stated in another way, if $(M,g)$ is non-stably causal then there
is an event $x\in M$ such that for every $\bar{g}>g$, $(M,\bar{g})$
is non-chronological at $x$.)
\end{lemma}

\begin{proof}
Let $(M,g)$ be non-stably causal and assume by contradiction that
for every $y\in M$ there is a ($y$ dependent) $\bar{g}_y>g$ such
that $(M,\bar{g}_y)$ is chronological at $y$. By Lemma \ref{pag},
taken $g_y$ such that $g< g_y<\bar{g}_y$, $(M,g_y)$ is strongly
causal at $y$ and hence it is strongly causal in an open
neighborhood $U_y$ of $y$ \cite{penrose72}.

Let $C$ be a compact. From the open covering $\{U_y, y \in C\}$, a
finite covering can be extracted $\{U_{y_1}, U_{y_2}, \dots,
U_{y_k}\}$. A metric $g_C>g$, on $M$ can be found such that for $
i=1,\ldots k$, $g_C<g_{y_i}$ on $M$. Thus $(M,g_C)$ is strongly
causal, and hence chronological, on a open set $A=\cup_i U_{y_i}
\supset C$. Let $(g_n,C_n, A_n)$ be a sequence of metrics $g_n>g$,
$g_{n+1}<g_{n}$, and strictly increasing compacts and open sets
$C_n\subset A_n \subset C_{n+1}$, such that $(M,g_n)$ is
chronological on $A_n$, and $\cup_n C_n=M$ (for instance introduce a
complete Riemannian metric and define $C_n$ as the balls of radius
$n$ centered at $x_0 \in M$, $C_n=B(x_0,n)$). Let $\chi_n:M \to
[0,1]$ be smooth functions such that $\chi_n=1$ on $C_n$, and
$\chi_n=0$ outside an open set $B_n$ such that
\[\cdots \subset C_n\subset B_n \subset \bar{B}_n\subset  A_n \subset
C_{n+1}\subset B_{n+1} \subset \bar{B}_{n+1} \subset  A_{n+1}
\subset \cdots.\] Construct a metric $g'>g$ on $M$ as follows (see
figure \ref{dim}). The metric $g'$ on $C_{n+1}\backslash B_n$ has
value $g_{n+1}$, and on $B_{n}\backslash C_{n}$ has value $\chi_n
g_n+(1-\chi_n) g_{n+1}$.

The spacetime $(M,g')$ is chronological otherwise there would be a
closed ($g'$-)timelike curve $\gamma$. Let $i$ be the minimum
integer such that $\bar{B}_i \cap \gamma \ne \emptyset$, and let $p
\in \bar{B}_i \cap \gamma$. Then $\gamma$ is also a closed
($g_i$-)timelike curve in $(M,g_i)$, thus chronology is violated at
$p \in \bar{B}_i \subset A_i$ a contradiction.

Thus $(M,g')$ is chronological, and hence $(M,g)$ is stably
chronological, or equivalently, stably causal (corollary \ref{pxe}).

\end{proof}

The next theorem was stated by Seifert \cite[Lemma 1]{seifert71}.
Unfortunately,  he did not give many details and I would say that
his argument can not be considered a proof.  Hawking and Sachs gave
another proof\footnote{There seems to be a gap in Hawking and
Sachs's proof. At the very beginning they state that given the
spacetime $(M,g)$ and $x\in M$, if $J^{+}_{S }$ is a partial order
then there is some $\bar{g}>g$ such that $(M,\bar{g})$ is causal at
$x$. However, they give no argument for this claim. It seems to me
that since $J^{+}_S$ is a partial order then for every $z \in M$,
there is a $\bar{g}_z>g$ such that $(M,\bar{g}_z)$ has no closed
causal curve which passes through $x$ and $z$, but, without a proof
of the contrary, $\bar{g}_z$ may well depend on $z$. Also note that
if the claim were obvious then lemma \ref{vax} would suffice to
prove the theorem. This gap is answered by lemma \ref{nhu}.} (see
the proof of \cite[Prop. 2.3]{hawking74}). The proof given here
differs from those and takes advantage of the previous lemmas.

\begin{theorem} \label{pai}
The relation $J^{+}_{S}$ on $M \times M$ is a partial order if and
only if $(M,g)$ is stably causal.
\end{theorem}

\begin{proof}
It is trivial that if $(M,g)$ is stably causal then $J^{+}_S$ is a
partial order. Indeed $(x,y) \in J^{+}_S$ and $(y,x) \in J^{+}_S$
imply that for a suitable $g'>g$, such that $(M,g')$ is causal,
$(x,y) \in J^{+}_{g'}$ and $(y,x) \in J^{+}_{g'}$, hence $x=y$
because of the causality of $(M,g')$.

For the converse let $J^{+}_{S}$ be a partial order, then for every
$x \in M$ there is  (lemma \ref{nhu}) a $x$-dependent metric $g_x>g$
such that $(M,g_x)$ is chronological at $x$, thus  $(M,g)$ is stably
causal because of lemma \ref{vax}.

\end{proof}

From corollary \ref{mkq} and theorem \ref{pai} it follows

\begin{corollary}
If $(M,g)$ is stably causal then it is $K$-causal.
\end{corollary}

\section{Violating sets on $M$ and $M\times M$} \label{vio}

In this section some results on the violating sets for $J^{+}_S$ are
obtained.

\begin{lemma} \label{mod}
The stable causality violating set on $M\times M$ for the spacetime
$(M,g)$ is the intersection of the chronological violating sets on
$M\times M$ for $g'>g$, namely
\begin{equation}
VJ_{S\, g}=\bigcap_{g'>g} VI_{g'},
\end{equation}
moreover $VJ_{S\, g}=\bigcap_{g'>g} VJ_{g'}=\bigcap_{g'>g}
V\bar{J}_{g'}=\bigcap_{g'>g} VK_{g'}$. Finally, $VI$ is open, while
$VJ_S\cup \Delta$,   $V \bar{J}\cup \Delta$ and  $VK\cup \Delta$ are
closed.
\end{lemma}

\begin{proof}
From lemma \ref{mip} $J^{+}_{S\, g}= \Delta\cup \bigcap_{g'>g}
I^{+}_{g'}$ thus
\[
J^{+}_{S\, g}\cap  J^{-}_{S\, g} \cap \Delta^{C}=(\bigcap_{g'>g}
I^{+}_{g'}) \cap (\bigcap_{g'>g} I^{-}_{g'}) \cap
\Delta^{C}=\bigcap_{g'>g} (I^{+}_{g'}\cap I^{-}_{g'}\cap
\Delta^{C}).
\]
The other equations are proved analogously, the last one using lemma
\ref{mip2}. It has been already mentioned that since $I^{+}$ is open
$VI$ is open. Since $VK\cup \Delta=K^+\cap K^{-}$ this set is closed
and an analogous argument holds for $V\bar{J}\cup \Delta$ and
$VJ_S\cup \Delta$.
\end{proof}

The original definition of stable causality implies that if $(M,g)$
is non-stably causal then for  every $g'>g$ there is a
($g$-dependent) event $x_{g'}\in M$ and a ($g'$-)timelike closed
curve through it. Actually, the equivalence between stable causality
and the property of antisymmetry for $J^{+}_S$, together with lemma
\ref{mod} imply a considerably stronger result

\begin{corollary}
If $(M,g)$ is non-stably causal then there is $(x,z)\in M\times M$,
$x\ne z$, such that for every $g'>g$, $(x,z) \in I^{+}_{g'}$ and
$(z,x) \in I^{+}_{g'}$.
\end{corollary}

\begin{lemma}
The stable causality violating set on $M$ for the spacetime $(M,g)$
is the intersection of the chronological violating sets on $M$ for
$g'>g$, namely
\begin{equation}
vJ_{S\, g}=\bigcap_{g'>g} vI_{g'},
\end{equation}
moreover $vJ_{S\, g}=\bigcap_{g'>g} vJ_{g'}=\bigcap_{g'>g}
v\bar{J}_{g'}=\bigcap_{g'>g} vK_{g'}$. Finally, $vI$ is open, while
$vJ_S$ and  $vK$ are closed (the proof of the closure of these sets
will follow from lemma \ref{loc}).
\end{lemma}

\begin{proof}
From  $vJ_{S\, g}=\pi(VJ_{S\, g})= \pi(\bigcap_{g'>g}
VI_{g'})\subset \bigcap_{g'>g} \pi(VI_{g'})=\bigcap_{g'>g} vI_{g'}$
an inclusion is obtained. The other direction follows by noticing
that if $x \in \bigcap_{g'>g} vI_{g'}$ then for every $g'>g$ there
is a ($g'$-)timelike closed curve passing through $x$. The proof of
lemma \ref{nhu} shows that under the same assumptions the existence
of an event $z\ne x$ can be inferred, such that $(x,z) \in
J^{+}_{S\, g}$ and $(z,x)\in J^{+}_{S \, g}$, that is $(x,z) \in
VJ_{S\, g}$, and finally $x \in vJ_{S\, g}$.

Projecting $VI \subset VJ \subset V\bar{J}\subset VK$ we obtain $vI
\subset vJ \subset v\bar{J}\subset vK$, which implies
$\bigcap_{g'>g} vI_{g'} \subset \bigcap_{g'>g} vJ_{g'}\subset
\bigcap_{g'>g} v\bar{J}_{g'}\subset \bigcap_{g'>g} vK_{g'}$. Recall
from lemma \ref{vfe2} that if $g <g'$ then $J^{+}_g\subset
\bar{J}^{+}_g\subset K^{+}_{g}\subset \Delta \cup I^{+}_{g'}$, thus
$VJ_g\subset V\bar{J}_g\subset VK_{g}\subset VI_{g'}$ and projecting
$vJ_g\subset v\bar{J}_g\subset vK_{g}\subset vI_{g'}$. We can now
prove all the other inclusions. I am going to prove that for every
$g'>g$, $\bigcap_{\bar{g}>g} vK_{\bar{g}} \subset vI_{g'}$, indeed I
can always find $\tilde{g}$, $g<\tilde{g}<g'$, and from
$\bigcap_{\bar{g} >g} vK_{\bar{g}} \subset vK_{\tilde{g}}\subset
vI_{g'}$ the thesis follows. Thus since for every $g'>g$,
$\bigcap_{\bar{g}>g} vK_{\bar{g}} \subset vI_{g'}$, it follows
$\bigcap_{\bar{g}>g} vK_{\bar{g}} \subset \bigcap_{{g}'>g} vI_{g'}$.
The equations are proved.

 It is well known
that $vI$ is open \cite[4.26]{penrose72}, that $vK$ is closed will
be proved in lemma \ref{loc} and from $vJ_{S\, g}=\bigcap_{g'>g}
vK_{g'}$ it follows the closure of $vJ_{S\, g}$.

\end{proof}

\begin{theorem}
$vI= M$ iff $I^{+}=M\times M$, analogously, $vJ_S=M$ iff
$J^{+}_S=M\times M$.
\end{theorem}

\begin{proof}
The first statement is well known, indeed $vI\ne M$ and $I^{+}\ne
M\times M$ are two equivalent definitions of a non-totally vicious
spacetime. The implication $I^{+}=M\times M \Rightarrow vI= M$ is
obvious, the other direction follows because $vI$ is made of
disjoint open components and for every $p,q$ in a same component
$p\ll q$. If $vI= M$ there is only one component which coincides
with $M$.

Let me prove the non-trivial part of the last statement, namely
$vJ_S=M$ $\Rightarrow J^{+}_S=M\times M$. Indeed,
$M=vJ_S=\bigcap_{g'>g}vI_{g'}$ $\Rightarrow  \forall g'>g, \
vI_{g'}=M$ $\Rightarrow I^{+}_{g'}=M\times M$ from the first
statement and hence $J^{+}_S= \bigcap_{g'>g} I^{+}_{g'}=M\times M$.

\end{proof}

\begin{remark} \label{nos}
Contrary to non-total viciousness which stays at the bottom of the
causal ladder, the condition $vJ_S \ne M$ has no place in the causal
hierarchy. Indeed, a causal spacetime may have $vJ_S = M$ (example
\ref{ex2} below), which may suggest that perhaps the property
$vJ_S\ne M$ is stronger than causality. However, it is easy to give
examples of non-chronological non-totally vicious spacetimes with
$vJ_S\ne M$ (identify two spacelike parallel lines in Minkowski
spacetime to obtain a cylinder and remove from it two parallel
spacelike half lines).

A good name for the condition $vJ_S\ne M$ is {\em stable non-total
viciousness}, because it holds iff $\exists g'>g: \ vI_{g'}\ne M$.
Nevertheless, it can also be {\em non-total non-stable causality},
because $vJ_S\ne \emptyset$ denotes non-stable causality, and
$vJ_S\ne M$ states that this non-stable causality is not total.
\end{remark}

\section{ $K$-causality and stable causality} \label{cex}

An important question is whether it is always $K^{+}=J^{+}_S$. The
answer is negative as the next examples prove

\begin{example} \label{ex1}
Consider the  1+1 cylindrical flat spacetime $M=\mathbb{R}\times
S^1$, of metric $\dd s^2=-\dd y\dd \theta$, $y \in \mathbb{R}$,
$\theta \in [0,2\pi)$. This spacetime is non-causal (hence
non-distinguishing) and reflecting, moreover given $x\in M$,
$K^{+}(x)=J^{+}(x)=\{z \in M: y(z)\ge y(x) \} \ne M$ while
$J^{+}_S(x)=M$ hence $K^{+}\ne J^{+}_S$.
\end{example}

\begin{example} \label{ex2}
The 2+1 spacetime $M=\mathbb{R}\times S^1 \times S^{1}$, of metric
$\dd s^2=-\dd y\dd \theta +\dd \phi^2$, $\theta,\phi \in
\mathbb{R}$, with the identifications
$(y,\theta,\phi)=(y,\theta,\phi+1)$ and
$(y,\theta,\phi)=(y,\theta+1,\phi+\alpha)$, $\alpha$ irrational
number is a causal non-distinguishing reflecting spacetime for which
given $x\in M$, $K^{+}(x)=\overline{J^{+}(x)}=\{z \in M: y(z)\ge
y(x) \} \ne M$ while $J^{+}_S(x)=M$ hence $K^{+}\ne J^{+}_S$. The
fact that $J^{+}_S=M$ follows for the same reason of the previous
example namely the compactness and lightlike nature of the space
section which in this case is a torus.
\end{example}

Although $K^{+}$ is not always coincident with $J^{+}_S$ it can be
that $K$-causality coincides with stable causality. For instance
this may happen because when the spacetime is $K$-causal the two
causal relations coincide as stated by claim \ref{c2}.

In order to proceed  we have to prove some statements regarding the
$K^{+}$ relation. First, recall that every event $x$ of the
spacetime $(M,g)$ admits an arbitrarily small convex neighborhoods
(and arbitrarily small globally hyperbolic neighborhoods). If $U$ is
such a neighborhood the causal relation on the spacetime $(U,g)$,
$J^{+}_U\subset U\times U$, is closed and hence coincides with the
relation $K^{+}_U$. This observation shows that if it were not for
global aspects the relation $K^{+}$ would be quite simple.

The next two lemmas were proved by Sorkin and Woolgar\footnote{The
version  given here  has slightly weaker assumptions because the set
$B$ is not required to be compact but only to have  compact
boundary. This difference will be important in the following.}
\cite[Lemmas 14,15]{sorkin96} using a {\em transfinite induction}
argument. Here I give  different proofs which use only soft
topological methods. I
 show that in most cases the transfinite induction argument can be
avoided. The suggested general strategy is as follows. First,
convert the property  to be proved into
 a causal relation on $M\times M$, then show that it is transitive,
closed and that contains $J^{+}$, finally use the minimality of
$K^{+}$. This approach is particularly clear if the statement of the
theorem is rearranged in a suitable way which reads ``Let $(x,z) \in
K^{+}$, if {\em hypothesis} then {\em thesis}''.

\begin{lemma} \label{sor}
Let $B\subset M$ be an open set of compact boundary $\dot{B}$. Let
$(x,z) \in K^{+}$, if $x \in B$ and $z \notin B$ (or viceversa),
then there is $y \in \dot{B}$ such that $(x,y) \in K^{+}$ and $(y,z)
\in K^{+}$.
\end{lemma}

\begin{proof}
Let $T^{+}\subset K^{+}$ be the set of pairs $(x,z) \in K^{+}$ at
which the statement of the theorem is true. This may happen for
instance because the hypothesis ``$x \in B$ and $z \notin B$ (or
viceversa)'' is false or because the hypothesis is true and the
thesis ``there is $y \in \dot{B}$ such that $(x,y) \in K^{+}$ and
$(y,z) \in K^{+}$'' is true.

It is $J^{+} \subset T^{+}$ because if $(x,z) \in J^{+}$ and the
hypothesis ``$x \in B$, $z \notin B$ (or viceversa)'' is true then
the thesis is true, $y$ being the intersection of the causal curve
$\sigma$ connecting $x$ to $z$ with $\dot{B}$ (the map $\sigma:
[0,1] \to M$ is defined over a compact, the set
$\sigma^{-1}(\bar{B})$ being closed and limited is a compact and
hence there is a last point $y$ at which the curve escapes
$\bar{B}$).

Also $T^{+}$ is closed, indeed if $(x,z) \in \bar{T}^{+}$ then
either ``$x \in B$ and $z \notin B$ (or viceversa)'' is false, in
which case $(x,z) \in T^{+}$ and there is nothing else to prove or
``$x \in B$ and $z \notin B$ (or viceversa)'' is true. Assume $x \in
B$ and $z \notin B$ the other case being analogous. There is a
sequence $(x_k,z_k)\in T^{+}\subset K^{+}$, $(x_k,z_k) \to (x,z)$,
and for sufficiently large $k$, $x_k \in B$. Now, if $z \in
\dot{B}$, then $(x,z) \in T^{+}$ because it satisfies the thesis of
the theorem with $y=z$. Thus we are left with the case $z \in
\bar{B}^{C}$ which is an open set, and hence for sufficiently large
$k$, $z_k \notin B$. Since $(x_k,z_k)\in T^{+}$, and the hypothesis
``$x_k \in B$ and $z_k \notin B$ (or viceversa)'' is satisfied,
there are $y_k \in \dot{B}$, $(x_k,y_k)\in K^{+}$ and $(y_k,z_k)\in
K^{+}$. Then there is an accumulation point $y \in \dot{B}$ and
since $K^{+}$ is closed, $(x,y) \in K^{+}$ and $(y,z) \in K^{+}$
which implies that $(x,z) \in T^{+}$ because the thesis of the
implication is true.

Finally, $T^{+}$ is transitive. Indeed, let $(x,w) \in T^{+}$ and
$(w,z) \in T^{+}$ then the only way in which $(x,z)$ could not
belong to $T^{+}$ is if $x\in B$, $z \notin B$ (or viceversa) and
the thesis is false. However, in this case $w$ must either belong to
$B$ or to $B^{C}$, in the former case since $(w,z) \in T^{+}$ there
must be the seeked $y \in \dot{B}$, $(x,w) \in K^{+}$, $(w,y)\in
K^{+}$ and $(y,z) \in K^{+}$ so that the thesis is verified because
$(x,y) \in K^{+}$. The latter case is analogous.

Thus $T^{+}\subset K^{+}$ is closed, transitive and contains
$J^{+}$. By the minimality of $K^{+}$ it is $T^{+}=K^{+}$ and hence
the implication of the theorem is true for every $(x,z) \in K^{+}$.

\end{proof}

Let $K^{+}_B$ the the $K^{+}$ relation for the spacetime $(B,g)$. It
can be regarded not only as a subset of $B\times B$ but also,
through the natural inclusion, as a subset of $M\times M$.


\begin{lemma} \label{sor2}
Let $B\subset M$ be an open set of compact boundary $\dot{B}$. Let
$(x,z) \in K^{+}$, if $x,z \in B$ and $(x,z) \notin K^{+}_B$ then
there is $y \in \dot{B}$ such that $(x,y) \in K^{+}$ and $(y,z) \in
K^{+}$.
\end{lemma}

\begin{proof}
Let $T^{+}\subset K^{+}$ be the set of pairs $(x,z) \in K^{+}$ at
which the statement of the theorem is true. This may happen for
instance because the hypothesis ``$x,z \in B$ and $(x,z) \notin
K^{+}_B$'' is false or because the hypothesis is true and the thesis
``there is $y \in \dot{B}$ such that $(x,y) \in K^{+}$ and $(y,z)
\in K^{+}$'' is true.

It is $J^{+} \subset T^{+}$ because if $(x,z) \in J^{+}$ and the
hypothesis `$x,z \in B$ and $(x,z) \notin K^{+}_B$'' is also true
then the thesis is true, $y$ being a point in the intersection
between the causal curve connecting $x$ to $z$ and $\dot{B}$. The
causal curve can not be entirely contained in $B$ otherwise $(x,z)
\in J^{+}_B\subset K^{+}_B$.

Also $T^{+}$ is closed, indeed if $(x,z) \in \bar{T}^{+}$  then
either ``$x,z \in B$ and $(x,z) \notin K^{+}_B$'' is false, in which
case $(x,z) \in T^{+}$ and there is nothing else to prove or ``$x,z
\in B$ and $(x,z) \notin K^{+}_B$'' is true. Let $x,z \in B$ and
$(x,z) \notin K^{+}_B$ and let $(x_k,z_k)\in T^{+}$,  be a sequence
such that $(x_k,z_k) \to (x,z)$. Since $B$ is open, for sufficiently
large $k$, $x_k,z_k \in B$, moreover we can assume $(x_k,z_k)\notin
K^{+}_B$. Indeed, if there is a subsequence $(x_s,z_s) \in K^{+}_B$
then $(x,z) \in K^{+}_B$ and hence $(x,z) \in T^{+}$ because the
hypothesis is false. Thus $x_k,z_k \in B$ and $(x_k,z_k)\notin
K^{+}_B$ and since $(x_k,z_k)\in T^{+}$ there are $y_k \in \dot{B}$,
such that $(x_k,y_k)\in K^{+}$ and $(y_k,z_k)\in K^{+}$. Then there
is an accumulation point $y \in \dot{B}$ and since $K^{+}$ is
closed, $(x,y) \in K^{+}$ and $(y,z) \in K^{+}$ which implies that
$(x,z) \in T^{+}$ because the thesis of the implication is true.

Finally, $T^{+}$ is transitive. Indeed, let $(x,w) \in T^{+}$ and
$(w,z) \in T^{+}$  then the only way in which $(x,z)$ could not
belong to $T^{+}$ is if $x,z\in B$, $(x,z) \notin K^{+}_B$, and the
thesis is false. However, in this case  $w$ must either belong to
$B$ or to $B^{C}$, in the former case  if  $(x,w) \in K^{+}_B$ and
$(w,z) \in K^{+}_B$ then $(x,z) \in K^{+}_B$ and hence $(x,z) \in
T^{+}$. If instead,  say $(x,w) \notin K^{+}_B$ (the other case
being analogous),  then since $(x,w) \in T^{+}$ and the thesis is
true for $(x,w)$ there is $y \in \dot{B}$, $(x,y) \in K^{+}$ and
$(y,w) \in K^{+}$ from which $(y,z) \in K^{+}$ and $(x,z) \in T^{+}$
follows. If instead $w \in B^{C}$, then lemma \ref{sor} can be
applied to $(x,w)$ to infer the existence of $y \in \dot{B}$ as
required by the thesis.

Thus $T^{+}\subset K^{+}$ is closed, transitive and contains
$J^{+}$. By the minimality of $K^{+}$ it is $T^{+}=K^{+}$ and hence
the implication of the theorem is true for every $(x,z) \in K^{+}$.

\end{proof}

The next result has been proved by Sorkin and Woolgar \cite[Lemma
16]{sorkin96} by making use of the previous lemmas and represents
the first step in Seifert's proof program. For completeness I
include the proof.


\begin{lemma} \label{sor3}
If $(M,g)$ is $K$-causal at $x$ then $x$ admits  arbitrarily small
$K$-convex neighborhoods. In particular, if $(M,g)$ is $K$-causal
then it is strongly $K$-causal.
\end{lemma}

\begin{proof}
Given $x \in M$ and $N\ni x$ a arbitrary neighborhood, there is
always  a strongly causal simple neighborhood $V\subset N$, $x \in
V$ (see, for instance, \cite[Sect. 2.3]{minguzzi06c}; recall that a
simple neighborhood is a convex neighborhood of compact closure
contained in another convex neighborhood \cite[Sect. 1]{penrose72}).
Since $V$ is convex $K^{+}_V=J^{+}_V$. Let $U_n\ni x$,
$\bar{U}_{n+1} \subset U_n$,  be a sequence of neighborhoods
causally convex (and hence $K$-convex) with respect to $V$. Let them
be a base for the topology at $x$, namely each open set containing
$x$ contains one $U_n$.

Assume there is a subsequence $U_k$ of non-$K$-convex neighborhoods.
There are $x_k, z_k \in U_k$, and $y_k \notin U_k$, such that
$(x_k,y_k) \in K^{+}$ and $(y_k,z_k) \in K^{+}$. The event $y_k$
belongs or not to $V$. In the former case it cannot be $(x_k,y_k)
\in K^{+}_V$ and $(y_k,z_k) \in K^{+}_V$ because $U_k$ is
$K_V$-convex, thus there is a pair among $(x_k,y_k)$ and $(y_k,z_k)$
to which lemma \ref{sor2} can be applied. The result is the
existence of $w_k \in \dot{V}$  such that $(x_k,w_k) \in K^{+}$ and
$(w_k,z_k) \in K^{+}$. In the latter case the application of lemma
\ref{sor} gives again the existence of $w_k \in \dot{V}$  such that
$(x_k,w_k) \in K^{+}$ and $(w_k,z_k) \in K^{+}$. Let $w \in \dot{B}$
be an accumulation point of the sequence $w_k$, then then since
$x_k,z_k \to x$, $(x,y) \in K^{+}$ and $(y,x) \in K^{+}$ in
contradiction with the $K$-causality at $x$. Thus for sufficiently
large $n$ all the open sets $U_n \subset N$ must be $K$-convex
neighborhoods.

\end{proof}

\begin{remark}
Note that $K$-causality implies strong $K$-causality instead of only
the strong $K_{g'}$-causality for $g'<g$ as one would expect from
analogy with
 lemma \ref{pag}. The reason lies in the fact that $K^{+}$ is
closed (in the proof of lemma \ref{pag} we could not infer $(c,d)
\in J^{+}_{g'}$ although $(c,d) \in \bar{J}^{+}_{g'}$, we had
instead to pass to $g>g'$ in order to close the causal chain).
\end{remark}

The next nice result is  due  to Sorkin and Woolgar \cite[Lemmas
12,13]{sorkin96}. The proof I give shows that one can use the
abstract notation $\circ$ for the composition so as to take
advantage of the distributive property with respect to unions of
sets.

\begin{lemma} \label{knj}
If $U$ is a open subset of $M$ then $K^{+}_U \subset
K^{+}\vert_{U\times U}$. Moreover, if $U$ is also $K$-convex then
$K^{+}_U = K^{+}\vert_{U\times U}$.
\end{lemma}

\begin{proof}

$K^{+}\vert_{U\times U}$ is closed (in the topology of $ U \times
U$), transitive and contains $J^{+}\vert_{U\times U}$, in
particular, $J^{+}_U \subset J^{+}\vert_{U\times U}$, thus because
of the minimality of $K^{+}_U$, $K^{+}_U\subset K^{+}\vert_{U\times
U}$.

The set $K^{+}_U$ can be regarded as a subset of $M\times M$ through
the natural inclusion. Consider the causal relation on $M$
\[
\tilde{K}^{+} = K^{+}_U \cup K^{+}\vert_{(U\times U)^{C}}.
\]
First note that $J^{+}_U\subset K^{+}_U$ and $J^{+} \vert_{(U\times
U)^{C}} \subset K^{+}\vert_{(U\times U)^{C}}$. Moreover, since
$K$-convexity implies causal convexity $J^{+}_U= J^{+}
\vert_{U\times U}$, thus $J^{+}\subset \tilde{K}^{+}$.

Since $K^{+}\vert_{(U\times U)^{C}}=K^{+}\cap (U\times U)^{C}$ and
$U$ is open,  this term is a closed set. The first term of
$\tilde{K}^{+}$, i.e. $K^{+}_U$, is closed in the topology of $U
\times U$ which is that induced from $M\times M$, thus it is closed
in $M\times M$, but, possibly, for accumulation points in
$\overline{K}^{+}_U\cap (U\times U)^{C}$, however
$\overline{K}^{+}_U \subset \bar{K}^{+}=K^{+}$ thus these points
belong to the second term and hence $\tilde{K}^{+}$ is closed.

Now, recall that $K^{+}_U$ is transitive, $K^{+}_U \circ K^{+}_U
\subset K^{+}_U$. The $K$-convexity of $U$ implies
$K^{+}\vert_{(U\times U)^{C}}\circ K^{+}\vert_{(U\times
U)^{C}}\subset K^{+}\vert_{(U\times U)^{C}}$. Moreover, $K^{+}_U
\circ K^{+}\vert_{(U\times U)^{C}}\subset K^{+}\vert_{(U\times
U)^{C}}$ because the first endpoint is not in $U$, and analogously
with the factors exchanged. The transitivity property for
$\tilde{K}^{+}$ is proved using the distributivity property of
$\circ$
\begin{align*}
 \tilde{K}^{+}\circ \tilde{K}^{+} &= [K^{+}_U\circ K^{+}_U] \cup
[K^{+}_U\circ K^{+}\vert_{(U\times U)^{C}}]\cup
[K^{+}\vert_{(U\times U)^{C}}\circ K^{+}_U]\\
&\quad \cup [K^{+}\vert_{(U\times U)^{C}}\circ K^{+}\vert_{(U\times
U)^{C}}]\subset\tilde{K}^{+}.
\end{align*}
Thus $\tilde{K}^{+}\subset K^{+}$, $J^{+}\subset \tilde{K}^{+}$ and
$\tilde{K}^{+}$ is transitive and closed  thus, because of the
minimality of $K^{+}$, $\tilde{K}^{+}=K^{+}$, and hence
$K^{+}\vert_{U\times U}=\tilde{K}^{+}\vert_{U\times U}=K^{+}_U$.

\end{proof}

\begin{lemma} \label{msd}
Let $(M,g)$ be a spacetime, and let $U$ be an open $K$-convex set of
compact closure contained in a convex open set $V$, then
$K$-causality holds at every point of $U$.
\end{lemma}

\begin{proof}

 Assume there is
${x} \in U$ at which $K$-causality does not hold, then there is ${z}
\in M$, ${z}\ne {x}$, such that $({x},{z}) \in K^{+}$ and $({z},{x})
\in K^{+}$. If ${z} \notin U$ then since ${z} \in K^{+}({x})\cap
K^{-}({x})$, $U$ would not be a $K$-convex. Thus ${z} \in U$.
Because of lemma \ref{knj}, it must be $({x},{z}) \in K^{+}_U$ and
$({z},{x}) \in K^{+}_U$
but this is impossible because
$J^{+}_V\vert_{U\times U}$ is closed (in the topology of $U \times
U$ as it is already closed in the topology of $V\times V$, the set
$V$ being convex \cite{oneill83}), transitive and contains
$J^{+}_U$, thus $K^{+}_U\subset J^{+}_V\vert_{U\times U}$ and hence
$({x},{z}) \in J^{+}_V$ and $({z},{x}) \in J^{+}_V$ in contradiction
with the causality of every convex neighborhood (it follows from the
fact that in a convex neighborhood  every pair of causally related
events is connected by a unique geodesic of well defined time
orientation, alternatively take $V$ inside a causal neighborhood).
\end{proof}

\begin{theorem} \label{loc}
The set $(vK)^{C} \subset M$ at which $(M,g)$ is $K$-causal is open.
\end{theorem}

\begin{proof}
Assume $(M,g)$ is $K$-causal at $x$, and let $V\ni x$ be a convex
set. The set $V$ exists and moreover $(V,g)$ is causal. Let $U$ be a
open $K$-convex set, $x \in U$,
 of compact closure contained in $V$. It exists because of lemma
\ref{sor3}. Thus by lemma \ref{msd} $K$-causality holds at every
point of $U$ and hence $(vK)^{C}$ is open.
\end{proof}

\begin{lemma} \label{kos}
If $g_1 \le g_2$ then $K^{+}_{g_1}\subset K^{+}_{g_2}$. In
particular, if $K_{g_2}$-causality holds at $x\in M$ then
$K_{g_1}$-causality holds at $x$.
\end{lemma}

\begin{proof}
$K^{+}_{g_2}\subset M\times M$ is (a) closed, (b) transitive and
contains $J^{+}_{g_2}\supset J^{+}_{g_1}$ and hence, (c) contains
$J^{+}_{g_1}$, thus must be larger than the smallest set which
satisfies (a), (b) and (c), namely $K^{+}_{g_1}$.
\end{proof}

\begin{lemma} \label{lsf}
Let $B\subset M$ be an open set of compact closure. Let
$N=M\backslash \bar{B}$, if  $x,z \in N$, $(x,z) \in K^{+}$ and
$(x,z) \notin K^{+}_N$ then there is $y \in \dot{B}$ such that
$(x,y) \in K^{+}$ and $(y,z) \in K^{+}$.
\end{lemma}

\begin{proof}
It is an immediate consequence of lemma \ref{sor2}, indeed $N$ is
open, is such that $M\backslash N$ is compact and plays the role of
the set $B$ in the statement of lemma \ref{sor2} (note that
$\dot{N}=\dot{B}$).
\end{proof}

\begin{lemma} \label{miy}
Let $B\subset M$ be an open set of compact closure. Let $\bar{g}\ge
g$, with $\{w\in M: \bar{g}(w)\ne g(w)\}\subset B$, let
$K_{g}$-causality hold at $x$, $x \notin \bar{B}$, then if
$K_{\bar{g}}$-causality does not hold at $x$ there is $z \in
\bar{B}$ at which $K_{\bar{g}}$-causality fails too.
\end{lemma}

\begin{proof}
Since $K_{\bar{g}}$-causality does not hold at $x$ there must be $z
\in M$, $z \ne x$, such that $(x,z) \in K^{+}_{\bar{g}}$ and $(z,x)
\in K^{+}_{\bar{g}}$. Assume $z\notin \bar{B}$ otherwise there is
nothing to prove. Since $K_{g}$-causality holds at $x$, $(x,z)
\notin K^{+}_{{g}}$ or $(z,x) \notin K^{+}_{{g}}$. Consider the
former case, the other being analogous. Let $N=M\backslash \bar{B}$,
from lemma \ref{knj} it is $K^{+}_{{g}}\vert_{N\times N} \supset
K^{+}_{(N,g)}=K^{+}_{(N,\bar{g})}$ so that $(x,z) \notin
K^{+}_{(N,\bar{g})}$. But it is also $(x,z) \in K^{+}_{\bar{g}}$ and
$x,z \in N$, thus by lemma \ref{lsf} there is $y \in \dot{B}$ such
that $(x,y) \in K_{\bar{g}}^{+}$ and $(y,z) \in K_{\bar{g}}^{+}$.
Composing these relations with $(z,x) \in K^{+}_{\bar{g}}$, it
follows $(x,y) \in K_{\bar{g}}^{+}$ and $(y,x) \in K^{+}_{\bar{g}}$
thus $K_{\bar{g}}$-causality does not hold at $y \in \bar{B}$.

\end{proof}

A fundamental observation is that given a causal relation
$R^{+}\subset M\times M$, the $R$-convexity of a set $A$ must be
understood as a condition on the shape of the metric on $A^{C}$
rather than on $A$. For instance, for $J$-convexity, the fact that
no causal curve can escape and reenter a causally convex set $A$ is
a constraint due to the shape of the light cones outside $A$. The
light cone structure inside $A$ has not very much to do with this
property. This observation is important because by enlarging the
light cones inside a $K$-convex set one expects to keep the
$K$-causality property. Since a special feature of $K$-causality is
that $K$-causality implies strong $K$-causality, the same
enlargement can be continued in other places so as to obtain, one
would say, a global widening of the light cones. This is basically
Seifert's program outlined by him in \cite[Lemma 1, point
(4)]{seifert71} (note that in that lemma ``at least once'' is
probably a misprint and must be replaced with ``at most once'').
Unfortunately, in order to follow this program, several technical
lemmas are needed. Some have been already proved. The next one is
particular because the lengthy proof works only if statements (a1),
(a2) and (b) are proved all at the same time.

\begin{lemma} \label{nbk}
Let $(M,g)$ be a spacetime,  $F$ a closed set, $B$ an open set of
compact boundary $\dot{B}$ and $F\subset B$. Let $(x,z) \in K^{+}$
\begin{itemize}
\item[(a1)] If $x \in \bar{B}$ and $z \notin \bar{B}$ then there is $y \in \dot{B}$ such
that $(x,y) \in K^{+}$ and $(y,z) \in K^{+}_{M\backslash F}$.
\item[(a2)] If $x \notin \bar{B}$ and $z \in \bar{B}$ then there is $y \in \dot{B}$ such
that $(x,y) \in K^{+}_{M\backslash F}$ and $(y,z) \in K^{+}$.
\item[(b)] If $x,z\notin \bar{B}$ then either $(x,z) \in K^{+}_{M\backslash F}$ or
there are $y_1,y_2 \in \dot{B}$ such that $(x,y_1) \in
K^{+}_{M\backslash F}$,  $(y_2,z) \in K^{+}_{M\backslash F}$,  and
$(y_1,y_2) \in K^{+}$.
\end{itemize}
In particular if $U$, $\bar{B}\subset U$, is an open set such that
$U\backslash F$ is $K_{M\backslash F}$-convex then $U$ is
$K$-convex.
\end{lemma}

\begin{proof}
Recall that $K^{+}_{M\backslash F} \subset K^{+}$. Let $T^{+}\subset
K^{+}$ be the set of pairs $(x,z) \in K^{+}$ at which all the three
statements of the theorem are true (which, selected a statement, may
happen because the hypothesis is false or because the thesis is
true).

It is $J^{+} \subset T^{+}$ because  the statements (a1), (a2) and
(b) are all true in this case. Indeed, consider (a1). If $(x,z) \in
J^{+}$ and the hypothesis ``$x \in \bar{B}$, $z \notin \bar{B}$'' is
also true then the thesis is true, $y$ being the last point of the
causal curve connecting $x$ to $z$ in $\bar{B}$ (the segment of the
causal curve connecting $y$ to $z$ is entirely contained in
$M\backslash F$).  The statement (a2) is proved similarly. As for
(b), if $(x,z) \in J^{+}$ and $x,z \notin \bar{B}$ then if the
causal curve connecting $x$ to $z$ does not intersect $\bar{B}$ then
$(x,z) \in J^{+}_{M\backslash F}$, otherwise there is a first point
$y_1$ at which the causal curve enters $\bar{B}$ and a last point
$y_2$ at which it leaves $\bar{B}$ so that the segments of causal
curves connecting $x$ to $y_1$ and $y_2$ to $z$ are contained on
$M\backslash F$ and hence $(x,y_1) \in J^{+}_{M\backslash F}$ and
$(y_2,z) \in J^{+}_{M\backslash F}$ while $(y_1,y_2) \in
J^{+}\subset K^{+}$ is obvious.

Also $T^{+}$ is closed, indeed if $(x,z) \in \bar{T}^{+}$ then
either $x,z \in \bar{B}$ in which case all the hypothesis of (a1),
(a2) and (b) are false, and thus the statements are true, $(x,z) \in
{T}^{+}$ and there is nothing left to prove, or only one of those
mutually excluding hypothesis is true.

Suppose the hypothesis of (a1) is true, that is, $x \in \bar{B}$,
$z\notin \bar{B}$. In this case statements (a2) and (b) are true
because their hypothesis are false and we have only to check that
statement (a1) is true. There is a sequence $(x_k,z_k)\in T^{+}$,
$(x_k,z_k) \to (x,z)$, and, since $\bar{B}^{C}$ is open, for
sufficiently large $k$,  $z_k \notin \bar{B}$. Now, without loss of
generality we can assume (pass to a subsequence if necessary) that
either (i) $x_k \in \bar{B}$ or (ii)  $x_k \notin \bar{B}$. In case
(i) since $(x_k,z_k)\in T^{+}$, and the hypothesis of (a1) ``$x_k
\in \bar{B}$ and $z_k \notin \bar{B}$'' is satisfied, there are $y_k
\in \dot{B}$, $(x_k,y_k)\in K^{+}$ and $(y_k,z_k)\in
K^{+}_{M\backslash F}$. Then there is an accumulation point $y \in
\dot{B}$ and since $K^{+}$ and $K^{+}_{M\backslash F}$ are both
closed, $(x,y) \in K^{+}$ and $(y,z) \in K^{+}_{M\backslash F}$
which implies that $(x,z) \in T^{+}$ because the thesis of (a1) and
hence statement (a1) is true. In case (ii) since $(x_k,z_k)\in
T^{+}$, and the hypothesis of (b) ``$x_k,z_k \notin \bar{B}$'' is
satisfied either there is a subsequence denoted in the same way such
that $(x_k,z_k) \in K^{+}_{M\backslash F}$ in which case $(x,z) \in
K^{+}_{M\backslash F}$ and the thesis of (a1) is verified with
$y=x$, or there are $y_{1k}, y_{2k} \in \dot{B}$, such that
$(x_k,y_{1k})\in K_{M\backslash F}^{+}$,  $(y_{2k},z_k)\in
K^{+}_{M\backslash F}$ and $(y_{1k},y_{2k}) \in K^{+}$. Then there
are accumulation points $y_1, y_2 \in \dot{B}$ and since
$K^{+}_{M\backslash F}$ and $K^{+}$ are closed, $(x,y_1) \in
K^{+}_{M\backslash F}$, $(y_2,z) \in K^{+}_{M\backslash F}$ and
$(y_1, y_2) \in K^{+}$. Thus $(x,y_2) \in K^{+}$ and $(y_2,z) \in
K^{+}_{M\backslash F}$ which implies that $(x,z) \in T^{+}$ because
the thesis of (a1) and hence statement (a1) is true along with (a2)
and (b).

The proof assuming true the hypothesis of (a2) is analogous.

Suppose the hypothesis of (b) is true, that is, $x,z \notin
\bar{B}$. In this case statements (a1) and (a2) are true because
their hypothesis is false and we have only to check that statement
(b) is true. There is a sequence $(x_k,z_k)\in T^{+}$, $(x_k,z_k)
\to (x,z)$, and, since  $\bar{B}^{C}$ is open, for sufficiently
large $k$, $x_k,z_k \notin \bar{B}$. Since $(x_k,z_k)\in T^{+}$ and
the hypothesis of (b) is true, for each (sufficiently large) $k$
either $(x_k,z_k) \in K^{+}_{M\backslash F}$ or there are
$y_{1k},y_{2k} \in \dot{B}$ such that $(x_k,y_{1k}) \in
K^{+}_{M\backslash F}$, $(y_{2k},z) \in K^{+}_{M\backslash F}$ and
$(y_{1k},y_{2k}) \in K^{+}$. If there is a subsequence such that the
first possibility holds then $(x,z) \in K^{+}_{M\backslash F}$ and
the thesis of (b) and hence statement (b) is true. Otherwise for all
but a finite number of values of $k$ the second possibility holds
then there are accumulation points $y_1$ of $y_{1k}$ and $y_2$ of
$y_{2k}$ so that, because of the closure of $K^{+}_{M\backslash F}$
and $K^{+}$, $(x,y_1) \in K^{+}_{M\backslash F}$, $(y_2,z) \in
K^{+}_{M\backslash F}$ and $(y_1, y_2) \in K^{+}$. Thus, again, (b)
is true because its thesis is true.

Finally, $T^{+}$ is transitive. Indeed, let $(x,w) \in T^{+}$ and
$(w,z) \in T^{+}$ then the only way in which $(x,z)$ could not
belong to $T^{+}$ is if $(x,z)$ contradicts one of the statements
(a1), (a2) or (b). Assume this happens for (a1) then $x \in \bar{B}$
and $z \notin \bar{B}$  while the thesis of (a1) is false (note that
(a2) and (b) are true because their hypothesis is false).
 However, in this case $w$ must either belong to
$\bar{B}$ or to $\bar{B}^{C}$, in the former case since $(w,z) \in
T^{+}$  and $w \in \bar{B}$, $z \notin \bar{B}$, there must be  $y
\in \dot{B}$, $(x,w) \in K^{+}$, $(w,y)\in K^{+}$ and $(y,z) \in
K_{M\backslash F}^{+}$ so that the thesis of (a1) is verified
because through composition $(x,y) \in K^{+}$. In the latter case
since $(x,w) \in T^{+}$  and $x \in \bar{B}$, $w \notin \bar{B}$,
there is $y \in \dot{B}$ such that $(x,y)\in K^{+}$ and $(y,w) \in
K^{+}_{M\backslash F}$. Unfortunately, the composition with $(w,z)
\in K^{+}$ can not be immediately done, however,  since $(w,z) \in
T^{+}$ and $w,z \notin \bar{B}$ either $(w,z) \in K_{M\backslash
F}^{+}$, and we have finished because $(x,y)\in K^{+}$ and $(y,z)
\in K_{M\backslash F}^{+}$ which is the thesis of (a1), or there is
$y_2\in \dot{B}$ (and an analogous $y_1$ of no interest here) such
that $(w,y_2) \in K^{+}$ and $(y_2,z)\in K^{+}_{M\backslash F}$.
Thus in this last case $(x,y_2) \in K^{+}$ and $(y_2,z) \in
K^{+}_{M\backslash F}$ that is, the thesis of (a1) is verified. Thus
statement (a1) can not be contradicted. The proof for (a2) is
analogous.

It remains to show that (b) can not be contradicted by $(x,z)$.
Assume $x,z \notin \bar{B}$, $w$ must either belong to $\bar{B}$ or
to $\bar{B}^{C}$ (note that (a1) and (a2) are true because their
hypothesis is false). In the former case since $(x,w) \in T^{+}$ and
$x \notin \bar{B}$, $w \in \bar{B}$, by (a2) there is $y_1 \in
\dot{B}$ such that $(x,y_1) \in K^{+}_{M\backslash F}$,  $(y_1,w)
\in K^{+}$. Also, since $(w,z) \in T^{+}$ and $w \in \bar{B}$, $z
\notin \bar{B}$ there is by (a1), $y_2 \in \dot{B}$ such that
$(w,y_2) \in K^{+}$ and $(y_2,z) \in K^{+}_{M\backslash F}$  thus
the thesis of (b) is true for $(x,z)$ as $(y_1,y_2) \in K^{+}$. It
remains to consider the case $w \notin \bar{B}$. In this case $(x,w)
\in T^{+}$, $(w,z) \in T^{+}$ and $x,w,z \notin \bar{B}$. There are
four possibilities depending on which of the cases given by the
thesis of (b) applies to the pairs $(x,w)$ and $(w,z)$. For
instance, by (b) it can be $(x,w) \in K^{+}_{M\backslash F}$. If
this is the case also for $(w,z)$ then $(x,z) \in K^{+}_{M\backslash
F}$ and the thesis of (b) holds for $(x,z)$ as required. Otherwise,
the second pair may satisfy the second alternative given by (b)
namely there are $y_1,y_2 \in \dot{B}$ such that $(w,y_1) \in
K^{+}_{M\backslash F}$, $(y_2,z) \in K^{+}_{M\backslash F}$ and
$(y_1,y_2) \in K^{+}$. Thus $(x,y_1) \in K^{+}_{M\backslash F}$,
$(y_2,z) \in K^{+}_{M\backslash F}$ and $(y_1,y_2) \in K^{+}$ which
again makes the thesis of (b) true for $(x,z)$. There are two cases
left.

The case $(x,w) \notin K^{+}_{M\backslash F}$, $(w,z) \in
K^{+}_{M\backslash F}$ is completely analogous to the one just
considered and leads again to the truth of (b). The last case $(x,w)
\notin K^{+}_{M\backslash F}$, $(w,z) \notin K^{+}_{M\backslash F}$
implies the existence of events $y_1, y_2, \bar{y_1}, \bar{y_2} \in
\dot{B}$ such that $(x,y_1) \in K^{+}_{M\backslash F}$, $(y_1,y_2)
\in K^{+}$, $(y_2,w) \in K^{+}_{M\backslash F}$, $(w,\bar{y}_1) \in
K^{+}_{M\backslash F}$, $(\bar{y}_1,\bar{y}_2) \in K^{+}$ and
$(\bar{y}_2,z) \in K^{+}_{M\backslash F}$. Thus setting
$\tilde{y}_1=y_1$, $\tilde{y}_2=\bar{y}_2$, it is $(x,\tilde{y}_1)
\in K^{+}_{M\backslash F}$, $(\tilde{y}_1,\tilde{y}_2) \in K^{+}$
and $(\tilde{y}_2,z) \in K^{+}_{M\backslash F}$, that is (b) is true
for $(x,z)$. Thus in every case $(x,z)$ can not contradict (b). The
transitivity of $T^{+}$ is proved.

Thus $T^{+}\subset K^{+}$ is closed, transitive and contains
$J^{+}$. By the minimality of $K^{+}$ it is $T^{+}=K^{+}$ and hence
the implications (a1), (a2) and (b) of the theorem are true for
every $(x,z) \in K^{+}$.

For the last statement of the theorem, if $U$ is not $K$-convex
there are $x,z \in U$, $y \notin U$ (thus $y\notin \bar{B}$) such
that $(x,y) \in K^{+}$ and $(y,z) \in K^{+}$. Thanks to (a1), (a2)
and (b) we can assume without loss of generality, through a
redefinition of $x$ and $z$, $(x,y) \in K_{M\backslash F}^{+}$ and
$(y,z) \in K_{M\backslash F}^{+}$ which contradicts the
$K_{M\backslash F}$-convexity of $U\backslash F$.
\end{proof}

\begin{lemma} \label{mum}
Let $F\subset M$ be closed set of compact boundary $\dot{F}$, and
$U\supset F$ an open set, then there is an open set $B$ of compact
boundary such that $F \subset B\subset \bar{B} \subset  U$.
\end{lemma}

\begin{proof}
For every point  $x \in \dot{F}$ it is possible to find a
neighborhood $A(x)\ni x$ of compact closure such that $\bar{A}(x)
\subset U$. Since $\dot{F}$ is compact there is a finite number of
points $x_i$, $i=1,\cdots, n$, such that $\{A(x_1),\cdots, A(x_n)\}$
covers $\dot{F}$. Then the open set $B=F\cup \bigcup_i A(x_i)$
 has compact boundary because $\dot{B}$ is a closed subset of the compact
 $\bigcup \dot{A}(x_1)$. Finally, by construction, $F
\subset B\subset \bar{B} \subset  U$.
\end{proof}

\begin{theorem} \label{mqp}
Let $U$ be an open subset of $M$ and let $F\subset U$ be a closed
set. If $U$ is $K$-convex then $U\backslash F$ is $K_{M\backslash
F}$-convex. Moreover, if $F$  has compact boundary  then the
converse holds, i.e. if $U\backslash F$ is $K_{M\backslash
F}$-convex then $U$ is $K$-convex.
\end{theorem}

\begin{proof}
Regard $K^{+}_{M\backslash F}$ as a subset of $M\times M$ through
the natural inclusion of $(M\backslash F) \times (M\backslash F)$
into $M\times M$. Assume $U$ is $K$-convex. From lemma \ref{knj},
$K^{+}_{M\backslash F}\subset K^{+}\vert_{(M\backslash F) \times
(M\backslash F)}$. Thus $U\backslash F$ is $K_{M\backslash
F}$-convex. For the converse, by lemma \ref{mum} there is a open set
$B$ of compact boundary $\dot{B}$, such that $F\subset B \subset
\bar{B} \subset U$. The conclusion comes from lemma \ref{nbk}.
\end{proof}

\begin{theorem} \label{mqr}
Let $(M,g)$ be a spacetime, $F$  a closed set of compact boundary,
and $U\supset F$ a open $K$-convex set. The spacetime is $K$-causal
iff $U$ is $K_U$-causal and $M\backslash F$ is $K_{M\backslash
F}$-causal.
\end{theorem}

\begin{proof}
Assume $(M,g)$ is $K$-causal then since with the usual natural
inclusion in $M\times M$, $K^{+}_U\subset K^{+}$, (lemma \ref{knj})
$U$ is $K_U$-causal and analogously since $K^{+}_{M\backslash F}
\subset K^{+}$, $M\backslash F$ is $K_{M\backslash F}$-causal.

For the converse, assume $M$ is not $K$-causal then there are $x,z
\in M$, $x \ne z$, such that $(x,z) \in K^{+}$ and $(z,x) \in
K^{+}$. If both are in $U$ then because of lemma \ref{knj} $(x,z)
\in K_U^{+}$ and $(z,x) \in K_U^{+}$ which is not possible since $U$
is $K_U$-causal by assumption. The case  $x \in U$, $z \notin U$
implies $z \in K^{+}(x,x)$ which contradicts the $K$-convexity of
$U$. The case $x \notin U$, $z \in U$ is analogous.

It remains the case $x,z \notin U$. By lemma \ref{mum} there is a
open set $B$ of compact boundary $\dot{B}$, such that $F\subset B
\subset \bar{B} \subset U$, thus  $x,z \notin \bar{B}$. Consider the
pair $(x,z) \in K^{+}$ and apply lemma \ref{nbk}. It follows that
either (i) $(x,z) \in K^{+}_{M\backslash F}$ or (ii) there are $y_1,
y_2 \in U\backslash F$ such that $(x,y_1) \in K^{+}_{M\backslash
F}$, $(y_1,y_2) \in K^{+}$ and $(y_2, z) \in K^{+}_{M\backslash F}$.
Consider case (ii) and $(z,x) \in K^{+}$. Applying again lemma
\ref{nbk} (iia) $(z,x) \in K^{+}_{M\backslash F}$ or (iib) there are
there are $\bar{y}_1, \bar{y}_2 \in U\backslash F$ such that
$(z,\bar{y}_1) \in K^{+}_{M\backslash F}$, $(\bar{y}_1,\bar{y}_2)
\in K^{+}$ and $(\bar{y}_2, x) \in K^{+}_{M\backslash F}$. But case
(iia) leads to $(y_2,z) \in K^{+}_{M\backslash F}$ and  $(z,y_1) \in
K^{+}_{M\backslash F}$ in contradiction with the $K_{M\backslash
F}$-convexity of $U\backslash F$. Analogously, case (iib) leads to
$(\bar{y}_2, x) \in K^{+}_{M\backslash F}$ and $(x,y_1) \in
K^{+}_{M\backslash F}$ which again contradicts the $K_{M\backslash
F}$-convexity of $U\backslash F$. Thus $(x,z) \in K^{+}_{M\backslash
F}$ and an analogous reasoning leads to $(z,x) \in
K^{+}_{M\backslash F}$ in contradiction with the $K_{M\backslash
F}$-causality of $M\backslash F$. The overall contradiction proves
that $M$ is $K$-causal.

\end{proof}

\begin{theorem} \label{mra}
Let $(M_1,g_1)$ and $(M_2,g_2)$ be two spacetimes  and let $F_1
\subset M_1$ and $F_2\subset M_2$ be two closed sets of compact
boundaries in the respective topologies. Define $N_1=M_1\backslash
\textrm{Int}\, F_1$, $N_2=M_2\backslash \textrm{Int} \,F_2$, and
assume there is an invertible isometry $\phi: N_1 \to N_2$ of the
spacetimes with boundaries $(N_1,g_1\vert_{N_1})$ and
$(N_2,g_2\vert_{N_2})$, $\phi(\dot{F}_1)=\dot{F}_2$.
Then open set $U_1 \supset F_1$ is $K_{(M_1,g_1)}$-convex iff
$U_2=\phi(U_1\backslash F_1)\cup F_2$ is $K_{(M_2,g_2)}$-convex.

Assume that $U_1$ is indeed $K_{(M_1,g_1)}$-convex,  and assume,
moreover, that $U_1$ is $K_{(U_1,g_1\vert_{U_1})}$-causal and $U_2$
is $K_{(U_2,g_2\vert_{U_2})}$-causal. Then $(M_1,g_1)$ is
$K_{(M_1,g_1)}$-causal iff $(M_2,g_2)$ is $K_{(M_2,g_2)}$-causal.
\end{theorem}

\begin{proof}

The set $U_1$ is $K_{(M_1,g_1)}$-convex iff $U_1\backslash F_1$ is
$K_{(M_1\backslash F_1,g_1\vert_{M_1\backslash F_1} )}$-convex
(theorem \ref{mqp}). The map $\phi\vert_{N_1\backslash \dot{F}_1}
:M_1\backslash F_1 \to M_2\backslash F_2$ is an isometry thus
$U_1\backslash F_1$ is $K_{(M_1\backslash
F_1,g_1\vert_{M_1\backslash F_1} )}$-convex iff $U_2\backslash
F_2=\phi(U_1\backslash F_1)$ is $K_{(M_2\backslash
F_2,g_2\vert_{M_2\backslash F_2} )}$-convex iff $U_2$ is
$K_{(M_2,g_2)}$-convex (theorem \ref{mqp}).

The last statement follows from theorem \ref{mqr} because, given the
assumptions, $(M_1,g_1)$ is $K_{(M_1,g_1)}$-causal iff
$(M_1\backslash F_1,g_1\vert_{M_1\backslash F_1})$ is
$K^{+}_{(M_1\backslash F_1,g_1\vert_{M_1\backslash F_1})}$-causal
which, due to the isometry, holds iff $(M_2\backslash
F_2,g_2\vert_{M_2\backslash F_2})$ is $K^{+}_{(M_2\backslash
F_2,g_2\vert_{M_2\backslash F_2})}$-causal which holds, again by
theorem \ref{mqr}, iff $(M_2,g_2)$ is $K_{(M_2,g_2)}$-causal.

\end{proof}

\begin{remark}
The theorem \ref{mra} is a  powerful result which allows to modify
the metric and even make surgery operations inside $U_1$. In short
it states that if a spacetime $(M_1,g_1)$ has a $K$-causal
$K$-convex open set $U_1$, given a closed set of compact boundary
$F_1 \subset U_1$  it is possible to arbitrary modify the metric and
even the topology inside $\textrm{Int}\, F_1$ without altering the
$K$-convexity of ``$U_1$'' in the sense that the obtained spacetime
$(M_2,g_2)$ will be such that (denoting with $C_1$ the complement in
$M_1$ and with $C_2$ the complement in $M_2$)
$U_2=((U_1)^{C_1})^{C_2}$ is $K$-convex.

Even more it states that whatever the metric deformation or surgery
operation done inside $U_1$, the spacetime does not lose its
$K$-causality provided the attached set $F_2$ does not make the
spacetime $(U_2,g_2\vert_{U_2})$ non-$K$-causal.

In the following we shall need to consider metric deformations
(theorem \ref{fgh}) and even topological surgery operations (theorem
\ref{las}).
\end{remark}

\begin{theorem} \label{fgh}
Let $(M,g)$ be a $K_g$-causal spacetime and let $x\in M$. Then a
metric $\bar{g}\ge g$, such that $\bar{g}>g$ in an open neighborhood
of $x$, exists such that  the spacetime $(M,\bar{g})$ is
$K_{\bar{g}}$-causal.
\end{theorem}

\begin{proof}

Since $(M,g)$ is $K$-causal it is also $K$-strongly causal. Thus it
is always possible to find  a open $K$-convex set $U\ni x$ and a
closed set of compact boundary $F$, $x \in \textrm{Int} F$, such
that $F\subset U$.  Even more $U$ can be chosen inside a  globally
hyperbolic neighborhood $V$. Since global hyperbolicity implies
stable causality and hence stable $K$-causality the metric can be
widened inside $F$ without spoiling the $K$-causality of $U$
(otherwise the $K$-causality of $V$ would be spoiled which would be
in contradiction with its stable $K$-causality). Finally, because of
theorem \ref{mra} the resulting spacetime is still $K$-causal.
%
%
%
%
\end{proof}

\begin{remark}
The previous result proves that the cones can be widened in a
neighborhood of any chosen point without spoiling $K$-causality.
Moreover, the only assumption of the theorem is $K$-causality itself
thus the procedure can be continued. The cones can be widened in a
finite number of points without spoiling $K$-causality. The problem
is that taken a point $y$ each of these enlargements may decrease
the size of the $K$-convex neighborhood around $y$. In particular
the kind of enlargement that can be applied at a neighborhood of a
new chosen point could depend on the sequence of enlargements
followed previously. Thus, though it would be natural to apply a
transfinite induction argument to assure the stable $K$-causality of
$(M,g)$, an argument in this direction would have to circumvent
these technical difficulties.
\end{remark}

\section{Some consequences of the possible coincidence} \label{pos}

Though the proof of the coincidence between $K$-causality and stable
causality has not been given, it is interesting to explore its
possible consequences. In this section I will clearly state if this
assumption is necessary for the results considered.

\begin{lemma} \label{cov}
Let $(M,g)$ be a spacetime. If $U_1$ and $U_2$ are $K$-convex sets
and $(U_1\times U_2) \cap (K^{+}\cup K^{-})=\emptyset$ then
$U=U_1\cup U_2$ is a $K$-convex set.
\end{lemma}

\begin{proof}
If $U$ is not $K$-convex there are $x,z \in U$ and $y \notin U$ such
that $(x,y) \in K^{+}$ and $(y,z) \in K^{+}$. Now, $x,z \in U_1$ is
excluded because $U_1$ is $K$-convex, analogously $x,z \in U_2$ is
excluded because $U_2$ is $K$-convex. Note that $(x,z) \in K^{+}$
(i.e. $(z,x) \in K^{-})$. The case $x \in U_1$, $z \in U_2$, is
excluded because $(U_1\times U_2) \cap K^{+}=\emptyset$ and
analogously the case $z \in U_1$, $x \in U_2$,  is excluded because
$(U_1\times U_2) \cap K^{-}=\emptyset$. Thus $U$ is $K$-convex.

\end{proof}

The next result comes from the development of an idea by H. Seifert
\cite[Lemma 2]{seifert71}.

\begin{theorem} \label{las}
Provided  $K$-causality coincides with stable causality: if $(M,g)$
is $K$-causal (stably causal) then $K^{+}=J^{+}_S$.
\end{theorem}

\begin{proof}
Assume by contradiction  $K^{+}_g\ne J^{+}_{S\, g}$ then  there is
$(x,z) \in J^{+}_{S\, g} \backslash K_g^{+}$, and in particular
$x\ne z$. It must also be $(z,x) \notin K^{+}_g$ otherwise $(z,x)
\in K^{+}_g \subset J^{+}_{S \, g}$, thus $J^{+}_{S \, g}$ would not
be a partial order an hence $(M,g)$ would not be stably causal, a
contradiction because $K$-causality and stable causality coincide.

We are going to construct a spacetime which is $K$-causal and yet
non-stably-causal in contradiction with the assumption.

 Since $K^{+}_g\cup K^{-}_g$ is
closed and $(x,z) \notin K^{+}_g\cup K^{-}_g$ there are  open sets
of compact closure $V_x \ni x$, and $V_z\ni z$ such that $(V_x
\times V_z)\cap (K^{+}_g\cup K^{-}_g)=\emptyset$.

Since $(M,g)$ is $K_g$-causal there is $\bar{g}>g$ such that $(M,
\bar{g})$  is $K_{\bar{g}}$-causal (equivalence between
$K$-causality, stable causality and stable $K$-causality, see
corollary \ref{pxe}) and hence strongly $K_{\bar{g}}$-causal. Let
$U_x \ni x$ (resp. $U_z \ni z$) be  a $K_{\bar{g}}$-convex
neighborhood contained in a ($\bar{g}$-)globally hyperbolic
neighborhood $W_x$ (resp. $W_z$ ) contained in $V_x$ (resp. $V_z$).
Let $u_x$ be a timelike vector at $x$, set \[D_x=\{y\in M: y=\exp_x
v, v \in TM_x, g(v,u_x)=0, 0 <g(v,v)<\epsilon\}\] where $\epsilon>0$
is chosen so that the closed disk (ball) $\bar{D}_x$ is contained in
$U_x$, and no ($\bar{g}$-)causal curve starting or ending at $x$ and
contained in $W_x$ can intersect $\bar{D}_x$ (this is possible
because $W_x$ being globally hyperbolic it is also distinguishing).
Let $D_z$ be defined analogously. Let $N=M\backslash(\bar{D}_x \cup
\bar{D}_z)$, then $(N, \bar{g}\vert_{N})$ is
$K^{+}_{(N,\bar{g}\vert_{N})}$-causal as
$K^{+}_{(N,\bar{g}\vert_{N})} \subset K^{+}_{(M,\bar{g})}$ (lemma
\ref{knj}). Cut $M$ at $\bar{D}_x$, that is let the boundary of $N$
be there topologically split into the two sides $D_x^{p}$ (above)
and $D_x^{f}$(below) (and analogously for $\bar{D}_z$).

Consider a globally hyperbolic spacetime $(G,g_2)$ with the topology
of $B^{n-1}\times (0,1)$ where $B^{n-1}$ is the open ball of
$\mathbb{R}^{n-1}$ ($n$  is the spacetime dimension). The metric on
it can be chosen so that its past boundary at $0$ can be glued with
$D^{f}_z$ and its future boundary at $1$ can be glued with $D^{p}_x$
all that preserving the continuity properties of the metric. Call
$\tilde{M}$ the final manifold and $\tilde{g}$ be the total metric
obtained joining $g\vert_N$ and $g_2$. The actual shape of the
metric $g_2$ is not important, but it must be chosen sufficiently
wide so that for every $z' \in D^{f}_z$, $D^{p}_x \subset
J^{+}_{(\tilde{M},\tilde{g})}(z')$.

I claim that the spacetime $(\tilde{M},\tilde{g})$ is
non-stably-causal. Indeed, we know that on $M$ for every $g'$,
$g<g'<\bar{g}$, $J^{+}_{S \, g} \subset \Delta \cup I^{+}_{g'}$
(lemma \ref{mip}) hence there is a ($g'$-)timelike curve connecting
$x$ to $z$. This curve can not intersect $D_x$ before escaping
$W_x$, nor can it intersect $D_z$ after entering
 $W_z$ (saved for the endpoints) because of the distinction of $W_x$, $W_z$, and  because of
the very definition of $D_x$ and $D_z$. However, it can not either
intersect $D_x$ after escaping $W_x$ because that would violate the
$K_{\bar{g}}$-convexity of $U_x$, $D_x \subset U_x \subset W_z$, and
analogously for $D_z$. Every metric $\tilde{g}'> \tilde{g}$ on
$\tilde{M}$ induces a metric $\tilde{g}'\vert_N>g\vert_{N}$ on $N$
which can be modified in a neighborhood of $D_x$ and $D_z$
 without changing the causal
nature of the curves connecting $x$ to $z$ so as to obtain an
extension  to a metric $g'$ of $M$. But since any metric $g'$ of
this form and such that $g'<\bar{g}$ allows for a ($g'$-)timelike
curve from $x$ to $z$, the same is true for for metrics $g'$ which
do not satisfy the constraint $g'<\bar{g}$. The conclusion is that
for every $\tilde{g}'>\tilde{g}$ there is a ($g'$-)causal and hence
($\tilde{g}'$-)causal curve connecting $x$ to $z$, and since $z$ and
$x$ are also connected by a $\tilde{g}$-causal curve passing through
$G$, the spacetime $(\tilde{M},\tilde{g})$ is not stably causal.

\begin{figure}
\centering \psfrag{a}{$D^p_x$} \psfrag{b}{$D^f_x$}
\psfrag{c}{$D^p_z$} \psfrag{d}{$D^f_z$} \psfrag{e}{$N$}
\psfrag{r}{$\dot{B}$} \psfrag{q}{$\dot{F}$} \psfrag{f}{$G$}
\psfrag{g}{$\!U_x$} \psfrag{h}{$U_z$} \psfrag{K}{$K^{+}(U_x)$}
\includegraphics[width=8cm]{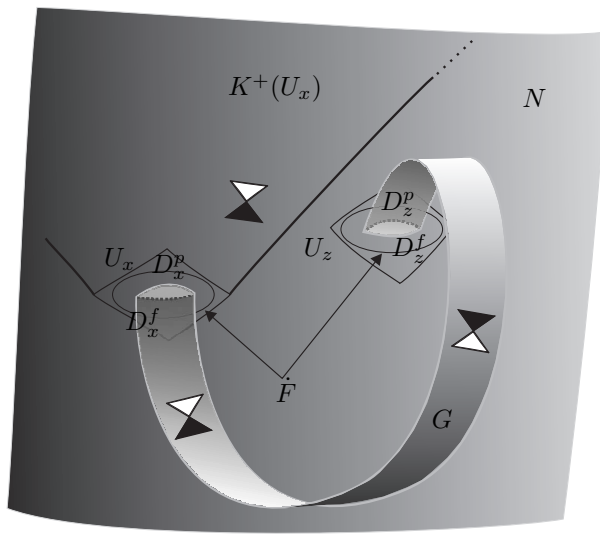}
\caption{The idea behind the proof of theorem \ref{las}. The dotted
disks $D^p_z$ and $D^f_x$ are boundaries of the total spacetime
$\tilde{M}=(M\backslash(\bar{D}_x \cup \bar{D}_z))\cup D^f_z \cup
G\cup D^p_x$.} \label{fig3}
\end{figure}

It remains to prove that $(\tilde{M},\tilde{g})$ is $K$-causal.

By lemma \ref{cov} $U=U_x \cup U_z$ is $K_{(M,g)}$-convex and
moreover, since $(M,g)$ is $K$-causal it is
$K_{(U,g\vert_U)}$-causal (theorem \ref{mqr}). The idea is to show
that $(\tilde{M}, \tilde{g})$ is obtained from $(M,g)$ through a
surgery operation allowed by theorem \ref{mra} which preserves
$K$-causality. Indeed, it is obvious that it is always possible to
find $F$, closed set of compact boundary, such that $(D_x\cup D_z)
\subset F\subset  U$. Then in this surgery operation
$\textrm{Int}(F)$ is excised (note that it is disconnected) and
replaced by a new set $\textrm{Int}(\tilde{F})$ (note that it is
connected) which is glued in place of it (all this made rigorous by
the obvious presence of the isometry cited in theorem \ref{mra}).
Intuitively, identifying some sets from $M$ and $\tilde{M}$,
$\textrm{Int}(\tilde{F})$ is the set $\textrm{Int}(\tilde{F})=D^f_z
\cup G\cup D^p_x\cup (\textrm{Int}(F)\backslash (\bar{D}_x \cup
\bar{D}_z)) $ while $\tilde{U}=D^f_z \cup G\cup D^p_x\cup
(U\backslash (\bar{D}_x \cup \bar{D}_z)) $. In the same sloppy
notation the boundaries of $F$ and $\tilde{F}$ can be identified.
Actually, note that because of the inclusion of $G$, while $F$ was
compact, the resulting $\tilde{F}$ is only closed but still of
compact boundary which is what is needed for applying theorem
\ref{mra}.  By theorem \ref{mra}, $\tilde{U}$
 is $K_{(\tilde{M},\tilde{g})}$-convex.
Moreover, $\tilde{U}$ is
$K_{(\tilde{U},\tilde{g}\vert_{\tilde{U}})}$-causal because  $G$ and
the metric on it can be chosen particularly well behaved.  In the
end all the conditions of theorem \ref{mra} are met and thus
$(\tilde{M},\tilde{g})$ is $K_{(\tilde{M},\tilde{g})}$-causal.

\end{proof}

A spacetime is reflecting if $(x,z) \in \bar{J}^{+}$ implies $z \in
\bar{J}^{+}(x)$ and $x \in \bar{J}^{-}(z)$ (this property can be
taken as a definition, see \cite[Prop. 3.45]{minguzzi06c}). Another
equivalent property is $z \in \bar{J}^{+}(x) \Leftrightarrow x \in
\bar{J}^{-}(z)$.

Recall that a spacetime is {\em causally continuous} if it is
reflecting and distinguishing.

If $K$-causality coincides with stable causality the next proof is
particularly simple. Compare for instance with the original one
given by Hawking and Sachs \cite{hawking74}.  Actually, it also
shows that the full assumptions underlying the definition of causal
continuity are not all needed for guaranteing  stable causality.

\begin{theorem} \label{pfs}
Assume that $K$-causality coincides with stable  causality. A
spacetime which is future distinguishing and future reflecting
 (or past distinguishing and past reflecting) is stably causal. In
 particular causal continuity implies stable causality.
\end{theorem}

\begin{proof}
It is a trivial consequence of \cite[Theor. 3.7]{minguzzi07b}
because the assumptions imply $K$-causality which is equivalent to
stable causality.
\end{proof}

The next result slightly  generalize previous results due to Dowker,
Garcia and Surya \cite[Prop. 2]{dowker00} and Hawking and Sachs
\cite[Theor. 2.1]{hawking74}. The proof that (i) and
distinction\footnote{  Examples of non-distinguishing spacetimes for
which (i) holds but (iv) does not hold are for instance examples
\ref{ex1} and \ref{ex2}.} implies (iv), originally given in
\cite[Theor. 2.1D]{hawking74}, would be  particularly simple
assuming the equivalence between stable causality and $K$-causality
in light of theorem \ref{las}.

\begin{theorem}
The following  conditions for the spacetime $(M,g)$ are equivalent
\begin{itemize}
\item[(i)] $(M,g)$ is future (resp. past) reflecting.
\item[(ii)] For every $x \in M$, $\uparrow\!I^{-}(x)=I^{+}(x)$ (resp.
$\downarrow\!I^{+}(x)=I^{-}(x)$).
\item[(iii)] For every $x \in M$, $\bar{J}^{+}(x)=K^{+}(x)$ (resp. $\bar{J}^{-}(x)=K^{-}(x)$).
\end{itemize}
Defined
\begin{itemize}
\item[(iv)] For every $x \in M$, $\bar{J}^{+}(x)=J_S^{+}(x)$ (resp. $\bar{J}^{-}(x)=J_S^{-}(x)$).
\end{itemize}
then $(iv) \Rightarrow (i)$, moreover if (i) holds in the past or
future case and the spacetime is  distinguishing  then (iv) holds (I
give the proof of the last statement assuming the equivalence
between $K$-causality and stable causality). Finally, if (i)-(iii)
hold in the past or future case, then $\bar{J}^{+}=K^{+}$, while if
(iv) holds in the past or future case then,
$\bar{J}^{+}=K^{+}=J^{+}_S$.
\end{theorem}

\begin{proof}
(i) $\Leftrightarrow$ (ii). The spacetime is future reflecting iff for every $x\in M$,
$A^{+}(x)=\bar{I}^{+}(x)$ that is iff $\overline{\uparrow\!I^{-}(x)}=\bar{I}^{+}(x)$,
that is iff ${\uparrow\!I^{-}(x)}={I}^{+}(x)$.  \\
(i) $\Rightarrow$ (iii). $\bar{J}^{+}$ is not only closed but also
transitive because if $(x,y) \in \bar{J}^{+}$ and $(y,z) \in
\bar{J}^{+}$ then \cite[Prop. 3.45]{minguzzi06c}, $y \in
\bar{J}^{+}(x)$ and $z \in \bar{J}^{+}(y)$ from which it follows
because of \cite[theorem 3.3]{minguzzi07b}  (or \cite[Claim
1]{dowker00}), $(x,z) \in \bar{J}^{+}$, hence $K^{+}=\bar{J}^{+}$.
In particular $\bar{J}^{+}(x)=\{y\in M :(x,y)\in
\bar{J}^{+}\}=\{y\in M :(x,y)\in K^{+}\}=K^{+}(x) $,
where in the first equality we used  the future reflectivity.  \\
(iii)$\Rightarrow$ (i). $x \in \bar{J}^{-}(z) \Rightarrow (x,z) \in
\bar{J}^{+} \Rightarrow (x,z) \in K^{+} \Leftrightarrow z \in
K^{+}(x)\Leftrightarrow  z \in \bar{J}^{+}(x)$, thus $(M,g)$ is future  reflecting. \\
(iv) $\Rightarrow$ (i). $x \in \bar{J}^{-}(z) \Rightarrow (x,z) \in
\bar{J}^{+} \Rightarrow (x,z) \in J_S^{+} \Leftrightarrow z \in
J_S^{+}(x)\Leftrightarrow  z \in \bar{J}^{+}(x)$, thus $(M,g)$ is future  reflecting. \\
Distinction and (i) $\Rightarrow$ (iv). I give the proof assuming
that $K$-causality coincides with stable causality. From (i), say
future case, it follows (iii) in the future case. Moreover, from
theorem \ref{pfs} the spacetime is stably causal ($K$-causal) thus,
by theorem \ref{las}, $K^{+}=J^{+}_S$, hence the thesis.

The last statement follows from the reflectivity since in this case
$\bar{J}^{+}=\{(x,z): z\in \bar{J}^{+}(x)\}$.
\end{proof}

\begin{remark} \label{jiu}
It can be $\bar{J}^{+}=K^{+}$ and yet the spacetime can be
non-reflecting. An example is provided by 1+1 Minkowski spacetime in
which a spacelike geodesic segment has been removed. Thus
$\bar{J}^{+}=K^{+}$ does not imply $(iii)$ whereas the converse
holds.
\end{remark}

\section{Conclusions}

The relationship between stable causality and $K$-causality and
their possible equivalence has been studied in detail. To this end
new results for the $K^{+}$ future have been obtained (lemma
\ref{nbk}, theorems \ref{mqp}, \ref{mqr}, \ref{mra}). Unfortunately,
a proof of the equivalence has not been given. A partial result in
the direction of the equivalence has been the proof that in a
$K$-causal spacetime, chosen an event, the light cones can be
widened in a neighborhood of the event without spoiling
$K$-causality (theorem \ref{fgh}). The process of enlarging the
light cones can be continued and thus, if the equivalence does
indeed hold, a final proof could be perhaps be obtained through an
inductive process starting from this result. In any case, if the
equivalence holds, in a $K$-causal spacetime the $K^+$ future
coincides with the Seifert future $J^{+}_S$ (theorem \ref{las}). If
the spacetime is not $K$-causal one expects to find some examples
which show that in general $K^+\ne J^{+}_S$ and indeed I gave the
example of a causal spacetime (example \ref{ex2}). Finally, a new
proof that Seifert's causal relation is a partial order iff the
spacetime is stably causal has been given which uses some new lemmas
which seem interesting in their own right (lemmas \ref{nhu} and
\ref{vax}).

\section*{Acknowledgements}
I warmly thank F. Dowker, R. Low, R.D. Sorkin and E. Woolgar, for
sharing  their impressions on the possible equivalence between
stable causality and $K$-causality.


\end{document}